%% file: baibao.tex
\documentclass[a4paper,12pt,reqno, oneside]{book}

\input seteps
\usepackage[pctex32]{graphics}
\usepackage {graphicx}
\usepackage{amsmath,amsxtra,amssymb,latexsym, amscd}
\usepackage[mathscr]{eucal}
\makeindex
\begin{document}
\parindent=1.05cm
\setlength{\baselineskip}{14truept}
\setcounter{page}{1}
\makeatletter
\title{\huge\bf
 Phenomenological models of baryon}
\author{{\bf Do Quoc Tuan}\\{\it Dept. of Theoretical Physics, College of Science, VNU, Hanoi, Vietnam}}
\date{}
\maketitle
\pagestyle{plain}

\pagestyle{myheadings}
\markboth{\footnotesize Phenomenological models of baryon }{\footnotesize  Do Quoc Tuan }

\vskip3cm
\begin{center}
{\large\bf
 Abstract
}
\end{center}
\vskip0.2cm
In this paper, I present almost my works performed during my time at VNU. I was interested in the composite Skyrme model proposed by H. Y. Cheung and F. Gursey. The expanding for this model based on results obtained from the original Skyrme model given by T. H. R. Skyrme in 1961 was general and interesting.

\newpage

\vskip-3cm
\begin{center}
{\large\bf
 Acknowledgments
}
\end{center}
\vskip0.2cm

I want to show deep gratitude to my supervisor Dr. Pham Thuc Tuyen (Department of Theoretical Physics, College of Science (another name is Hanoi University of Science), Vietnam National University) because of everything that he has been teaching and supporting me during a student's time!
\vskip0.2cm
I also thank Department of Theoretical Physics and Faculty of Physics for giving me the best conditions to completing the thesis!
\vskip0.2cm
Finally, and most of all, I want to give my thesis to my parent, my beloved as a special gift for the supports over a long period!
\begin{flushright}
Ha noi, 6/2007

Do Quoc Tuan\footnote{email: do.tocxoan@gmail.com}
\end{flushright}
\tableofcontents
\leftskip0cm
\chapter{Why is the Skyrme model}
\section{The weak point of QCD}
\vskip0.2cm
In physically, we can divide the nature in two types of particles: matter particles and field particles. Matter particles interact together by fields concluding transfer interaction particles. There are four fields corresponding four interactions (four forces) and their particles. The graviton is a transfer interaction particle of gravitational force, the photon is a transfer interaction particle of electromagnetic force, $W^ \pm  $ and $Z^o $ are transfer interaction particle of weak force, the gluon is transfer interaction particle of strong force. It is known well, quarks are interact together by the strong force. In other word, the interaction of quarks is the exchange of gluons. And the theory that describes quarks and their interaction is $QCD$ ({\bf Quantum Chromo Dynamics}). In mathematically, Gell-Mann and Neuman proposed that quarks make up a multiple which represents the gauge $SU(3)$ being the group of $3 \times 3$ unitary matrices with determinant {\bf 1}. Gell-Mann firstly gives generators of $SU(3)$, therefore, they are called Gell-Mann matrices $\lambda $ . They satisfy the Lie algebra
\begin{equation}
\left[ {\lambda _i ,\lambda _j } \right] = 2if_{ijk} \lambda _k ,
\end{equation}
where $f_{ijk} $ is the structure constant. It is totally antisymmetric
\begin{equation}
f_{ijk}  =  - f_{jik} .
\end{equation}
\vskip0.2cm

To describe the interaction between quarks, we need using the gluon field $A_\mu ^\alpha  $ . And now, the Lagrangian density is the Yang-Mills Lagrangian density
\begin{equation}
{\cal L} =  - \frac{1}{4}F_{\mu \nu }^\alpha  F^{\mu \nu \alpha }  + i\bar \psi \gamma ^\mu  (\partial _\mu   + igA_\mu  )\psi  + M\bar \psi \psi ,
\end{equation}
\begin{equation}
A_\mu = A_\mu ^\alpha  \lambda ^\alpha ,
\end{equation}
\begin{equation}
F_{\mu \nu }^\alpha   = \partial _\mu  A_\nu ^\alpha   - \partial _\nu  A_\mu ^\alpha   - gf^{\alpha \beta \gamma } A_\mu ^\beta  A_\nu ^\gamma ,
\end{equation}
where {\bf g} is a constant of interaction. At the high energy scale ($>TeV$), $QCD$ is very good to use. But, when the energy lower (in $GeV$), it meets with some problems in which describe quantities of quarks and gluons. The main cause is the high value of {\bf g}, it obstructs an expansion of interaction terms of hadron. To repair this problem, t'Hooft proposed that $1/N_c$ plays as {\bf g}($N_c$ is color number in a gauge theory). And, he gave that in large-$N_c$ limit, $QCD$ becomes equivalent to an effective field theory of meson.
\section{The idea of Skyrme}
\vskip0.2cm

In 1961, Skyrme gave the new model in which baryons may be obtained from the theory of meson [1]. He began with the Lagrangian density of pion
\begin{equation}
{\cal L} = \frac{1}{2}\partial _\mu  \vec \pi \partial ^\mu  \vec \pi  - \frac{{m^2 }}{2}\vec \pi \vec \pi ,
\end{equation}
in which  $\pi$'s make up a three dimensional isotopic space. After, he expanded the isotopic $SU\left( 2 \right)$ becomes a new group with the algebra
\begin{equation}
\left[ {V_i ,V_j } \right] = 2i\varepsilon _{ijk} V_k ,
\end{equation}
\begin{equation}
\left[ {A_i ,A_j } \right] = 2i\varepsilon _{ijk} V_k ,
\end{equation}
\begin{equation}
\left[ {V_i ,A_j } \right] = 2i\varepsilon _{ijk} A_k ,
\end{equation}
where generators $V_i $ are vector charges, generators $A_i $ are axial charges. And now, an isotopic space have four components $\left\{ {\vec \pi } \right\} \to \left\{ {\vec \varphi ,\varphi _4 } \right\}$ transforming in a rule
\begin{equation}
\left[ {V_i ,\varphi _j } \right] = 2i\varepsilon _{ijk} \varphi _k ,
\end{equation}
\begin{equation}
\left[ {V_i ,\varphi _4 } \right] = 0 ,
\end{equation}
\begin{equation}
\left[ {A_i ,\varphi _j } \right] =  - 2i\varepsilon _{ij} \varphi _4 ,
\end{equation}
\begin{equation}
\left[ {A_i ,\varphi _4 } \right] = 2i\varphi _i .
\end{equation}
\vskip0.2cm

Particles described by $\varphi _i \left( {i = 1,2,3} \right)$ are pseudo- scalars and making isotopic vectors. A particle described by $\varphi _4 $ has a positive parity, a spin and an isotopic spin equals zero, and it is a vacuum.
\vskip0.2cm
If replace $V_i $ and $A_i $ by $L_i $ and $R_i $ defined
\begin{equation}
L_i  = \frac{1}{2}\left( {V_i  - A_i } \right) ,
\end{equation}
\begin{equation}
R_i  = \frac{1}{2}\left( {V_i  + A_i } \right) ,
\end{equation}
then obtaining a new algebra

\begin{center}
$\left[ {L_i ,L_j } \right] = \left[ {\frac{1}{2}\left( {V_i  - A_i } \right),\frac{1}{2}\left( {V_j  - A_j } \right)} \right]$ ,
\end{center}
\begin{center}
$= \frac{1}{4}\left[ {V_i ,V_j } \right] - \frac{1}{4}\left[ {V_i ,A_j } \right] - \frac{1}{4}\left[ {A_i ,V_j } \right] + \frac{1}{4}\left[ {A_i ,A_j } \right]$ ,
\end{center}
\begin{center}
$= \frac{1}{2}i\varepsilon _{ijk} V_k  - \frac{1}{2}i\varepsilon _{ijk} A_k  + \frac{1}{2}i\varepsilon _{jik} A_k  + \frac{1}{2}i\varepsilon _{ijk} V_k $ ,
\end{center}
\begin{center}
$= i\varepsilon _{ijk} V_k  - i\varepsilon _{jik} A_k $ ,
\end{center}
\begin{equation}
\bf{\Rightarrow \left[ {L_i ,L_j } \right] = 2i\varepsilon _{ijk} L_k}
\end{equation}
and
\begin{center}
$\left[ {R_i ,R_j } \right] = \left[ {\frac{1}{2}\left( {V_i  + A_i } \right),\frac{1}{2}\left( {V_j  + A_j } \right)} \right]$ ,
\end{center}
\begin{center}
$= \frac{1}{4}\left[ {V_i ,V_j } \right] + \frac{1}{4}\left[ {V_i ,A_j } \right] + \frac{1}{4}\left[ {A_i ,V_j } \right] + \frac{1}{4}\left[ {A_i ,A_j } \right]$ ,
\end{center}
\begin{center}
$= \frac{1}{2}i\varepsilon _{ijk} V_k  + \frac{1}{2}i\varepsilon _{ijk} A_k  - \frac{1}{2}i\varepsilon _{jik} A_k  + \frac{1}{2}i\varepsilon _{ijk} V_k $ ,
\end{center}
\begin{center}
$= i\varepsilon _{ijk} V_k  + i\varepsilon _{jik} A_k $ ,
\end{center}
\begin{equation}
\bf{\Rightarrow \left[ {R_i ,R_j } \right] = 2i\varepsilon _{ijk} R_k}
\end{equation}
and
\begin{center}
$\left[ {L_i ,R_j } \right] = \left[ {\frac{1}{2}\left( {V_i  - A_i } \right),\frac{1}{2}\left( {V_j  + A_j } \right)} \right]$ ,
\end{center}
\begin{center}
$= \frac{1}{4}\left[ {V_i ,V_j } \right] + \frac{1}{4}\left[ {V_i ,A_j } \right] - \frac{1}{4}\left[ {A_i ,V_j } \right] - \frac{1}{4}\left[ {A_i ,A_j } \right]$ ,
\end{center}
\begin{center}
$= \frac{1}{2}i\varepsilon _{ijk} V_k  + \frac{1}{2}i\varepsilon _{ijk} A_k  + \frac{1}{2}i\varepsilon _{jik} A_k  - \frac{1}{2}i\varepsilon _{ijk} V_k $ ,
\end{center}
\begin{equation}
\bf{\Rightarrow \left[ {L_i ,R_j } \right] = 0} .
\end{equation}
\vskip0.2cm

This is an algebra of the chiral group $SU\left( 2 \right) \times SU\left( 2 \right)$. Thus, the expanding group is the chiral group $SU\left( 2 \right) \times SU\left( 2 \right)$. Continuously, he proposed that $\left\{ {\vec \varphi ,\varphi _4 } \right\}$ form a three dimensional sphere
\begin{equation}
\vec \varphi ^2  + \varphi _4^2  = 1 .
\end{equation}
\vskip0.2cm
Recently, $\left\{ {\vec \varphi ,\varphi _4 } \right\}$ are associated with $2 \times 2$ unitary $U\left( {x,t} \right)$ matrix
\begin{center}
$U\left( {x,t} \right) = \exp i\vec \tau \vec \pi \left( {x,t} \right)$ ,
\end{center}
\begin{center}
$= \exp i\vec \tau \frac{{\vec \pi }}{\pi }\pi$ ,
\end{center}
\begin{equation}
= \cos \pi  + i\vec \tau \frac{{\vec \pi }}{\pi }\sin \pi ,
\end{equation}
where $\pi  = \sqrt {\vec \pi ^2 } $, $\tau $'s are Pauli matrices. Pion fields now are associated with $\vec \pi $ fields
\begin{equation}
\varphi _4  \Leftrightarrow \cos \pi ,
\end{equation}
\begin{equation}
\vec \varphi  \Leftrightarrow \frac{{\vec \pi }}{\pi }\sin \pi .
\end{equation}
\vskip0.2cm
Really
\begin{equation}
\vec \varphi ^2  + \varphi _4^2  = \frac{{\vec \pi ^2 }}{{\pi ^2 }}\sin ^2 \pi  + \cos ^2 \pi  = 1 .
\end{equation}
\vskip0.2cm

The Lagrangian density proposed by Skyrme [1] is
\begin{equation}
{\cal L} = \frac{{F_\pi ^2 }}{{16}}Tr\left( {\partial _\mu  U\partial _\mu  U^\dag  } \right) + \frac{1}{{32e^2 }}Tr\left[ {\left( {\partial _\mu  U} \right)U^\dag  ,\left( {\partial _\nu  U} \right)U^\dag  } \right]^2 ,
\end{equation}
here $U$ is an $SU\left( 2 \right)$ matrix transforming as $U \to AUB^{ - 1}$ under an chiral $SU\left( 2 \right) \times SU\left( 2 \right)$. He proved that the model has a soliton solution (called hedgehog\footnote{firstly called by Polyakov, see at {\bf L. Ryder}, {\it Quantum Field Theory} (second edition), Cambridge University Press} solution)
\begin{equation}
U_o  = \exp \left( {i\vec \tau \frac{{\vec r}}{r}F\left( r \right)} \right) ,
\end{equation}
where $F\left( r \right)$ is the chiral angle, baryons are obtained from the quantization of soliton solution(see chapter 2).
\vskip0.2cm

It is  well known that in the large-$N$ limit, $QCD$ may
be regarded as the theory of effective meson fields. E. Witten argued
that baryons may be regarded as solitons of this effective meson
theory without an further reference to their quark
content [2]. We know that solitons have not a spin. But, by
quantizing around solitons, we obtain the special topological
construct of these solitons which makes a spin. On the other hand,
soliton solutions have a limited spatial distribution, therefore,
they are correlative to particles that have a limited spatial
dimension [2,15].

\begin{figure}[htb]
\centering
\includegraphics[scale=0.6]{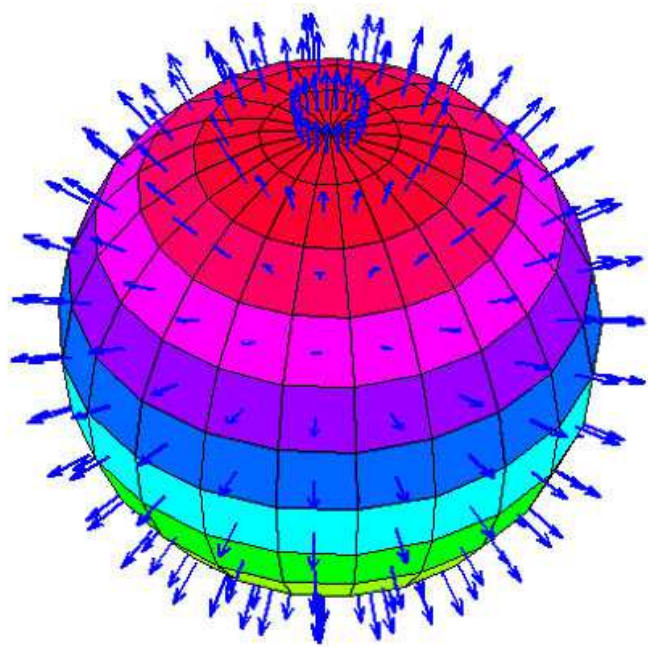}
\caption{The configuration of hedgehog, arrows on it are isotopic vectors of $\vec c$}
\end{figure}
\chapter{The massless composite Skyrme model}
\section{The formalism}

In 1983, G. S. Adkins, C. R. Nappi and E. Witten used the
Skyrme's idea to computed the static properties of baryons on the
chiral group $SU\left( 2 \right) \times SU\left( 2 \right)$ and
found that the model gives predictions that agree well with
experiments [2]. However, this model has a problem in which the
situation of particles at low energy are not interested but it
regards that particles in free states, and this point can reduce the
exactitude of theoretical datum. In 1989, this problem is treated by
P. T. Tuyen and N. A. Viet [3]. They considered that at low
energy particles are not free but they would shrink together into
the cluster. Therefore, the group that describes them is $SU\left( 2
\right) \times SU\left( 2 \right)/SU\left( 2 \right)_{diag} $. This
model gives few datum better than the sigma model. In 1990, H. Y.
Cheung, F. Gursey [4] gave the general model in which the
sigma model and the {\bf V-T} model are two special cases corresponding
with $n=1$ and $n=2$. They introduced a composite \emph{SU(2)}
configuration $V_n = UPUP...PU$, where \emph{U(x,t)} is the
\emph{SU(2)} matrix and $P = p_o  + i\vec \tau .\vec p$ is the
constant \emph{SU(2)} matrix, and they proposed the composite Skyrme
lagrangian density
\begin{equation}
{\cal L}_n  =  - \frac{{F_\pi ^2 }}{{16n^2 }}Tr\left( {\partial _\mu
V^{ - n} \partial ^\mu  V^n } \right) + \frac{1}{{32e^2 n^4
}}Tr\left( {\left[ {V^{ - n} \partial _\mu  V^n ,V^{ - n} \partial
_\nu  V^n } \right]^2 } \right) ,
\end{equation}
with $F_\pi  $ is the pion decay constant, \emph{\textbf{e}} is a
dimensionless parameter.
\vskip0.2cm

Under the chiral group $SU\left( 2 \right)
\times SU\left( 2 \right)$, \emph{U} transforms as $U \to AUB^\dag
$, \emph{P} transforms as $P \to BPA^\dag  $, and so $V_n $
transforms as $V_n  \to AV_n B^\dag  $. In the limit $P \to I$, $V_n
\to U_n $, the lagrangian (2.1) becomes approximated by the lagrangian
density
\begin{equation}
{\cal L}_n  =  - \frac{{F_\pi ^2 }}{{16n^2 }}Tr\left( {\partial _\mu
U^{ - n} \partial ^\mu  U^n } \right) + \frac{1}{{32e^2 n^4
}}Tr\left( {\left[ {U^{ - n} \partial _\mu  U^n ,U^{ - n} \partial
_\nu  U^n } \right]^2 } \right) .
\end{equation}
\vskip0.2cm

Now, using the hedgehog soliton solution
gave by Skyrme $U_0 \left( r \right) = \exp \left[ {i\tau .\hat
rF\left( r \right)} \right]$, where $\tau $'s are Pauli's matrices
and $\hat r = \vec r/r$, the lagrangian density (2.2) is
\begin{equation}
{\cal L}_n  = \frac{{e^2 F_\pi ^4 }}{8}\left[ {\left(
{\frac{{dF}}{{d\tilde r}}} \right)^2  + 2\frac{{\sin ^2 nF}}{{n^2
\tilde r^2 }}} \right] + \frac{{e^2 F_\pi ^4 \sin ^2 nF}}{{2n^2
\tilde r^2 }}\left[ {2\left( {\frac{{dF}}{{d\tilde r}}} \right)^2
+ \frac{{\sin ^2 nF}}{{n^2 \tilde r^2 }}} \right],
\end{equation}
where $\tilde r = eF_\pi  r$ is the dimensionless variable.
\vskip0.2cm And the static energy of hedgehog is
\begin{equation}
{\cal E}_n  = \frac{{F_\pi  }}{e}A_n  = M_n ,
\end{equation}
\begin{equation}
A_n  = \int\limits_0^\infty  {4\pi \tilde r^2 \left\{
{\frac{1}{8}\left[ {\left( {\frac{{dF}}{{d\tilde r}}} \right)^2  +
2\frac{{\sin ^2 nF}}{{n^2 \tilde r^2 }}} \right] + \frac{{e^2
F_\pi ^4 \sin ^2 nF}}{{2n^2 \tilde r^2 }}\left[ {2\left(
{\frac{{dF}}{{d\tilde r}}} \right)^2  + \frac{{\sin ^2 nF}}{{n^2
\tilde r^2 }}} \right]} \right\}} d\tilde r.
\end{equation}
\vskip0.2cm

We obtain the non-linear differential equation of
\emph{F(r)} from minimum condition of the hedgehog's energy
\begin{equation}
\delta _F {\cal E}_n =0 ,
\end{equation}
or
\begin{equation}
\frac{d}{{dr}}\frac{{\delta {\cal E}_n}}{{\delta F' }} - \frac{{\delta {\cal E}_n}}{{\delta F}} = 0 , ( Euler-Lagrange-equation)
\end{equation}
\begin{equation}
\left( {\frac{{\tilde r^2}}{4} + \frac{2}{{n^2 }}\sin ^2 nF}
\right)F''  + \frac{{\tilde r}}{2}F'  + \frac{1}{n}\sin 2nF\left(
{F'^2  - \frac{{\sin ^2 nF}}{{n^2 \tilde r^2 }} - \frac{1}{4}}
\right) = 0 .
\end{equation}
\vskip0.2cm
To solve this equation we need boundary conditions for the
\emph{F(r)}. The first condition of \emph{F(r)} that relates to
the limit of hedgehog's energy when $r \to \infty $ is $F\left(
\infty \right) = 0$. The second condition relating to the exist of
baryonic number B is $F\left( 0 \right) = \pi $. This equation can
not give us the analytic solution. So we will solve it by the
numerical method.
\begin{figure}[htb]
\centering
\includegraphics[scale=0.5]{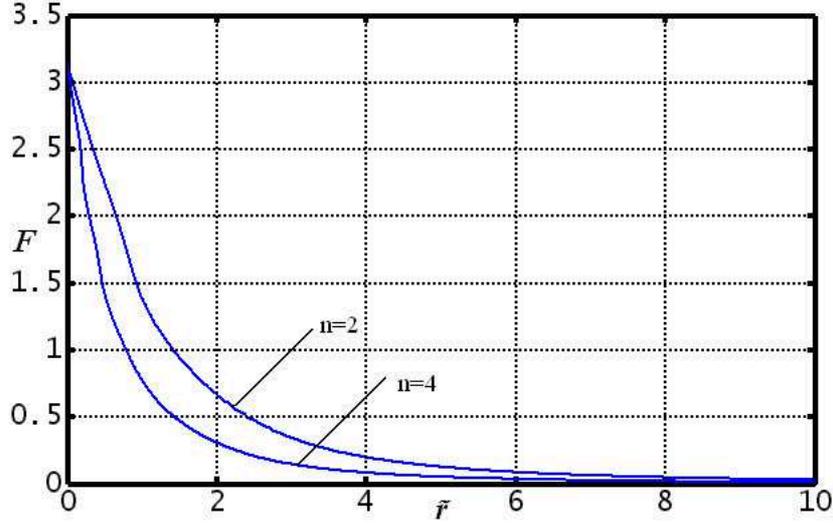}
\caption{The numerical solution of (2.8) for cases of n=2 and n=4 [22, 25].}
\end{figure}
\vskip0.2cm

Now we will quantize the soliton by collective
coordinates $A\left( t \right) = a_o \left( t \right)+ i\vec
a\left( t \right)\vec \tau $  with  $a_\mu  \left( t \right)$ $\left( {\mu  = 0,1,2,3} \right)$ are collective coordinates, $A\left( t \right)$ is $SU(2)$ matrix. From properties of $SU(2)$ matrix, we have
\begin{center}
$A\left( t \right)A^\dag  \left( t \right) = 1$ ,
\end{center}
\begin{center}
$\left\{ {a_o \left( t \right) + i\vec a\left( t \right)\vec \tau } \right\}\left\{ {a_o \left( t \right) - i\vec a\left( t \right)\vec \tau } \right\} = 1$ ,
\end{center}
\begin{equation}
\bf{a_\mu  a_\mu = 1} .
\end{equation}
By time derivative, resulting
\begin{equation}
\bf{a_\mu  \dot a_\mu = 0}
\end{equation}

By this way, solitons to be particles. Different quantization take
different particles. Thus, it takes us to the new idea about the
origin of particles in nature in which they are made by a basic
particle (called soliton). With the \emph{A(t)} matrix, the $U^n
\left( {x,t} \right)$ will transform as \[ U^n \left( {x,t}
\right) = A\left( t \right)U_0^n \left( x \right)A^{ - 1} \left( t
\right)
\], and we put it into (2.2), the lagrangian now is
\begin{center}
$L_n  =  - {\cal E}_n  + \Gamma _n Tr\left[ {\partial _0 A\left( t
\right)\partial _0 A^{ - 1} \left( t \right)} \right]$ ,
\end{center}
\begin{equation}
= - {\cal E}_n  +
2\Gamma _n \sum\limits_{\mu  = 0}^3 {\dot a_\mu ^2 } ,
\end{equation}
with
\begin{equation}
\Gamma _n  = \frac{1}{{e^3 F_\pi  }}\Phi _n ,
\end{equation}
\begin{equation}
\Phi _n  = \frac{{2\pi }}{{3n^2 }}\int\limits_0^\infty  {\tilde r^2
\sin ^2 nF\left\{ {1 + 4\left[ {\left( {\frac{{dF}}{{d\tilde r}}}
\right)^2  + \frac{{\sin ^2 nF}}{{n^2 \tilde r^2 }}} \right]}
\right\}} d\tilde r.
\end{equation}
\vskip0.2cm

The conjugate momenta are obtained from the derivative
of lagrangian following $\dot a_\mu(t)$
\begin{center}
$\pi _\mu   = \frac{{\partial L_n }}{{\partial \dot a_\mu  }}$ ,
\end{center}
\begin{center}
$=\frac{{\partial \left[ { - M_n  + 2\Gamma _n \sum\limits_{\mu  = 0}^3 {\dot a_\mu ^2 } } \right]}}{{\partial \dot a_\mu }}$ ,
\end{center}
\begin{equation}
 = 4\Gamma _n \dot a_\mu  .
\end{equation}
\vskip0.2cm

We go to the hamiltonian
\begin{center}
$H_n  = \pi _\mu  \dot a_\mu   - L_n $ ,
\end{center}
\begin{center}
$= 4\Gamma _n \dot a_\mu  \dot a_\mu - L_n $ ,
\end{center}
\begin{center}
$= M_n  + 2\Gamma _n \dot a_\mu  \dot a_\mu $ ,
\end{center}
\begin{equation}
= M_n  + \frac{{\left( {\sum\limits_{\mu  = 0}^3 {\pi _\mu ^2 } } \right)}}{{8\Gamma _n }} .
\end{equation}
\vskip0.2cm

Canonical quantizing momenta $\pi _\mu = - i\frac{\partial }{{\partial a_\mu }}$ let the hamiltonian as
\begin{equation}
H_n  = M_n  + \frac{1}{{8\Gamma _n }}\sum\limits_{\mu  = 0}^3 {\left( {\frac{{ - \partial ^2 }}{{\partial a_\mu ^2 }}} \right)} .
\end{equation}
\vskip0.2cm

Now, we will define the isotopic vector  $\vec c$  following the way
\vskip0.1cm
\begin{center}
$\dot A^\dag  A = \left( {\dot a_0  - i\dot a_i \tau _i }
\right)\left( {a_0  + ia_i \tau _i } \right) $ ,
\end{center}
\vskip0.1cm
\begin{center}
$= \dot a_0 a_0 + \dot \vec a\vec a - i\left( {a_0 \dot \vec a - \dot a_0 \vec a +
\vec a \times \dot \vec a} \right)\vec \tau $ ,
\end{center}
note (2.10), we have
\begin{equation}
\dot A^\dag  A=- i\vec c\vec \tau  ,
\end{equation}
with
\begin{equation}
\vec c = \left( {a_0 \dot \vec a - \dot a_0 \vec a + \vec a \times \dot \vec a} \right) .
\end{equation}
\vskip0.2cm The square isotopic vector $\vec c$  is
\begin{center}
\vspace{0.1cm}
$\vec c^2  = \dot a_0^2 \vec a^2  + \dot \vec a^2 a_0^2  + \vec a^2 \dot \vec a^2  - (\vec a\dot \vec a)^2  - 2\dot a_0 a_0 \dot \vec a\vec a$ ,
\end{center}
\begin{center}
\vspace{0.1cm}
$= \dot a_0^2 \vec a^2  + \dot \vec a^2 a_0^2  + \vec a^2 \dot \vec a^2  - (\vec a\dot \vec a)^2  + 2(\vec a\dot \vec a)^2$ ,
\end{center}
\begin{equation}
= \dot a_\mu  \dot a_\mu  .
\end{equation}
\vskip0.2cm

In the configuration of hedgehog, since the isotopic
vector and the radius vector have the same direction, rotation in
the co-ordinate space can make an isospin of particle, and we
assume that a spin equals to an isospin. The canonical momenta
that correspond with collective coordinates are isospins. We can
homogenize $\vec c^2 $ with the square of a spin and an isospin
\begin{equation}
\vec c^2 = \frac{1}{{16\Gamma _n^2 }}\vec J^2 = \frac{1}{{16\Gamma
_n^2 }}\vec T^2 .
\end{equation}
\vskip0.2cm Thus, we have
\begin{equation}
H_n  = M_n  + \frac{1}{{8\Gamma _n }}\vec J^2  = M_n  +
\frac{1}{{8\Gamma _n }}\vec T^2 ,
\end{equation}
with $\vec J$ is a spin and $\vec T$ is an isospin. Eigenvalues of
hamiltonian (2.21) are
\begin{equation}
E_n  = M_n  + \frac{{l\left( {l + 2} \right)}}{{8\Gamma _n }} ,
\end{equation}
with \emph{l=2J} and \emph{J} is the quantum number of spin. So,
the nucleon $\left( {J = 1/2} \right)$ and the delta $\left( {J = 3/2}
\right)$ masses are given by
\begin{equation}
M_N = M_n  + \frac{3}{{8\Gamma _n }}  ,
\end{equation}
\begin{equation}
M_\Delta = M_n  +
\frac{{15}}{{8\Gamma _n }} .
\end{equation}
\vskip0.2cm
Note (2.12), (2.13), (2.23), (2.24) we have
\begin{equation}
F_\pi = \sqrt[4]{{\frac{{\left( {5M_N  - M_\Delta  } \right)^3
\left( {M_\Delta   - M_N } \right)\Phi _n }}{{384A_n^3 }}}}  ,
\end{equation}
\begin{equation}
e =\frac{{4A_n F_\pi  }}{{\left( {5M_N  - M_\Delta  } \right)}} .
\end{equation}

\section{Calculations for static quantities of baryon }
\vskip0.2cm After constructing the formalism of massless composite model, we apply to the baryon (nucleon). Since this model is constructed depend on some ideas of theory of topology then we can start at the topological current to calculations for static quantities of baryon [2, 3, 4, 15].
\vskip0.2cm The topological current (a baryonic current) is
defined
\begin{equation}
B^\mu   = \frac{{\varepsilon ^{\mu \nu \alpha \beta } }}{{24\pi ^2 n}}Tr\left[ {\left( {U^{\dag n} \partial _\nu  U^n } \right)\left( {U^{\dag n} \partial _\alpha  U^n } \right)\left( {U^{\dag n} \partial _\beta  U^n } \right)} \right],
\end{equation}
with $\varepsilon _{0123}  =  - \varepsilon ^{0123}  = 1$ (totally antisymmetry).
\vskip0.2cm If we put $U^n \left( {x,t} \right) = A\left( t
\right)U_0^n \left( x \right)A^{ - 1} \left( t \right)$ into (2.27),
we will have
\begin{equation}
B^0  =  - \frac{1}{{2\pi ^2 }}\frac{{\sin ^2 nF}}{{r^2 }}F' ,
\end{equation}
\begin{equation}
B^i  = \frac{{i\varepsilon ^{{\rm{ijk}}} }}{{2\pi ^2 }}\frac{{\sin
^2 nF}}{{r^2 }}F'x_k Tr\left[ {\dot A^\dag  A\tau _j } \right] .
\end{equation}
\vskip0.2cm Note (2.17) and (2.20), then (2.29) is now
\begin{equation}
B^i  = \frac{{\varepsilon ^{{\rm{ijk}}} }}{{2\pi ^2 }}\frac{{\sin
^2 nF}}{{r^2 }}F'x_k \frac{{J_j }}{{2\Gamma _n }} .
\end{equation}
\vskip0.2cm The baryonic number is
\begin{equation}
B = \int\limits_0^\infty  {4\pi r^2 } B^0 \left( r \right)dr = 1 .
\end{equation}
\vskip0.2cm The iso-scalar electric square radius is
\begin{equation}
\left\langle {r^2 } \right\rangle _{E,I = 0}  =
\int\limits_0^\infty  {4\pi r^2 } B^0 \left( r \right)r^2 dr .
\end{equation}
\vskip0.2cm The iso-scalar magnetic square radius is
\begin{equation}
\left\langle {r^2 } \right\rangle _{M,I = 0}  =
\frac{3}{5}\frac{{\int\limits_0^\infty  {4\pi r^2 B^0 \left( r
\right)r^4 } }}{{\left\langle {r^2 } \right\rangle _{E,I = 0} }} .
\end{equation}
\vskip0.2cm We have the time's component of
vector current $\vec V$
\begin{equation}
V_i^0  = \frac{i}{3}\frac{{\sin ^2 nF}}{{n^2 r^2 }}\left( {F_\pi
^2  + \frac{4}{{e^2 }}\left[ {\left( {\frac{{dF}}{{dr}}} \right)^2
+ \frac{{\sin ^2 nF}}{{n^2 r^2 }}} \right]} \right)Tr\left( {\dot
AA^\dag  \tau _i } \right) ,
\end{equation}
note (2.17)
\begin{equation}
V_i^0  = \frac{{\sin ^2 nF}}{{3n^2 r^2 }}\left( {F_\pi ^2  +
\frac{4}{{e^2 }}\left[ {\left( {\frac{{dF}}{{dr}}} \right)^2  +
\frac{{\sin ^2 nF}}{{n^2 r^2 }}} \right]} \right)\frac{{T_i
}}{{2\Gamma _n }} .
\end{equation}
\vskip0.2cm The density of charge of nucleon is defined by
Gell-Mann - Nishijima
\begin{equation}
Q = \int\limits_0^\infty  {\rho _{nucleon} } \left( r \right)dr =
I_3  + \frac{1}{2}B ,
\end{equation}
and
\begin{equation}
I_3  = \int\limits_0^\infty  {r^2 } V_3^0 d\Omega dr =
\int\limits_0^\infty  {\frac{{4\pi \sin ^2 nF}}{{3n^2 }}\left(
{F_\pi ^2  + \frac{4}{{e^2 }}\left[ {\left( {\frac{{dF}}{{dr}}}
\right)^2  + \frac{{\sin ^2 nF}}{{n^2 r^2 }}} \right]}
\right)\frac{{T_3 }}{{2\Gamma _n }}} dr .
\end{equation}
\vskip0.2cm So, we have
\begin{equation}
\rho _{nu} \left( r \right) = \frac{{4\pi \sin ^2 nF}}{{3n^2
}}\left( {F_\pi ^2  + \frac{4}{{e^2 }}\left[ {\left(
{\frac{{dF}}{{dr}}} \right)^2  + \frac{{\sin ^2 nF}}{{n^2 r^2 }}}
\right]} \right)\frac{{T_3 }}{{2\Gamma _n }} - \frac{{\sin ^2
nF}}{\pi }F' .
\end{equation}
With $\bf {T_3  = \frac{1}{2}}$ and $\bf {T_3  =  - \frac{1}{2}}$, we have the
charge's densities of proton and neutron.
\vskip0.2cm

After integrating, we obtain charges of a proton and a neutron
\begin{equation}
Q_{pro}  = \int\limits_0^\infty  {\rho _{pro} } \left( r \right)dr = 1 ,
\end{equation}
\begin{equation}
Q_{neu}  = \int\limits_0^\infty  {\rho _{neu} } \left( r \right)dr = 0 .
\end{equation}
\vskip0.2cm
The iso-vector electric square radius and the iso-vector magnetic square radius are
\begin{equation}
\left\langle {r^2 } \right\rangle _{M,I = 1}  = \left\langle {r^2 } \right\rangle _{E,I = 1}  = \int\limits_0^\infty  {r^2 V_3^0 } dr = \infty .
\end{equation}
\vskip0.2cm
However, in the experiment the iso-vector electric radius is {\bf 0.88}, the iso-vector magnetic radius is {\bf 0.8}. This is the weak point of the massless composite Skyrme model!
\vskip0.2cm
The iso-scalar magnetic moment is defined
\begin{equation}
\vec \mu _{I = 0}  = \frac{1}{2}\int {\left( {\vec r \times \vec
B} \right)d^3 x} ,
\end{equation}
and the result is
\begin{equation}
\vec \mu _{I = 0}  = \frac{1}{{3\Gamma _n }}\left\langle {r^2 }
\right\rangle _{I = 0} \vec J .
\end{equation}
\vskip0.1cm
The connection between
the spin and the magnetic moment is
\begin{equation}
\vec \mu  = \frac{g}{{2M_N }}\vec J .
\end{equation}
\vskip0.1cm Note (2.23, 2.24), we have
\begin{equation}
g_{I = 0}  = g_{proton}  + g_{neutron}  = \frac{4}{9}M_N \left(
{M_\Delta   - M_N } \right)\left\langle {r^2 } \right\rangle _{I =
0} .
\end{equation}
\vskip0.1cm The iso-vector magnetic moment is defined
\begin{equation}
\vec \mu _{I = 1}  = \frac{1}{2}\int {\left( {\vec r \times \vec
V_3 } \right)d^3 } x ,
\end{equation}
where $\vec V_3 $ is the vector current. After few calculations,
the result is
\begin{equation}
\vec \mu _{I = 1}  = \frac{1}{3}\Gamma _n \vec J .
\end{equation}
\vskip0.1cm And we also have
\begin{equation}
g_{I = 1}  = g_{proton}  - g_{neutron}  = \frac{{2M_N }}{{M_\Delta
- M_N }} .
\end{equation}
\vskip0.1cm Using (2.44) and (2.47), we can find
\begin{equation}
\mu _{pro}  = \frac{{g_{pro} }}{2} = \frac{1}{4}\left[
{\frac{4}{9}M_N \left( {M_\Delta   - M_N } \right)\left\langle
{r^2 } \right\rangle _{I = 0}  + \frac{{2M_N }}{{M_\Delta   - M_N
}}} \right] ,
\end{equation}
\begin{equation}
\mu _{neu}  = \frac{{g_{neu} }}{2} = \frac{1}{4}\left[
{\frac{4}{9}M_N \left( {M_\Delta   - M_N } \right)\left\langle
{r^2 } \right\rangle _{I = 0}  - \frac{{2M_N }}{{M_\Delta   - M_N
}}} \right] .
\end{equation}

\vskip0.2cm

The axial coupling is defined
\begin{equation}
g_A  =  - \frac{\pi }{{3e^2 }}D_n  ,
\end{equation}
with
\begin{equation}
D_n  = \int\limits_0^\infty  {\tilde r^2 } \left\{ {\frac{{\sin
2nF}}{{n\tilde r}}\left[ {1 + 4\left( {F'^2  + \frac{{\sin ^2
nF}}{{n^2 \tilde r^2 }}} \right)} \right] + F'\left( {1 +
\frac{{8\sin ^2 nF}}{{n^2 \tilde r^2 }}} \right)} \right\}d\tilde
r .
\end{equation}
\newpage
\begin{figure}[htb]
\centering
\includegraphics[scale=0.40]{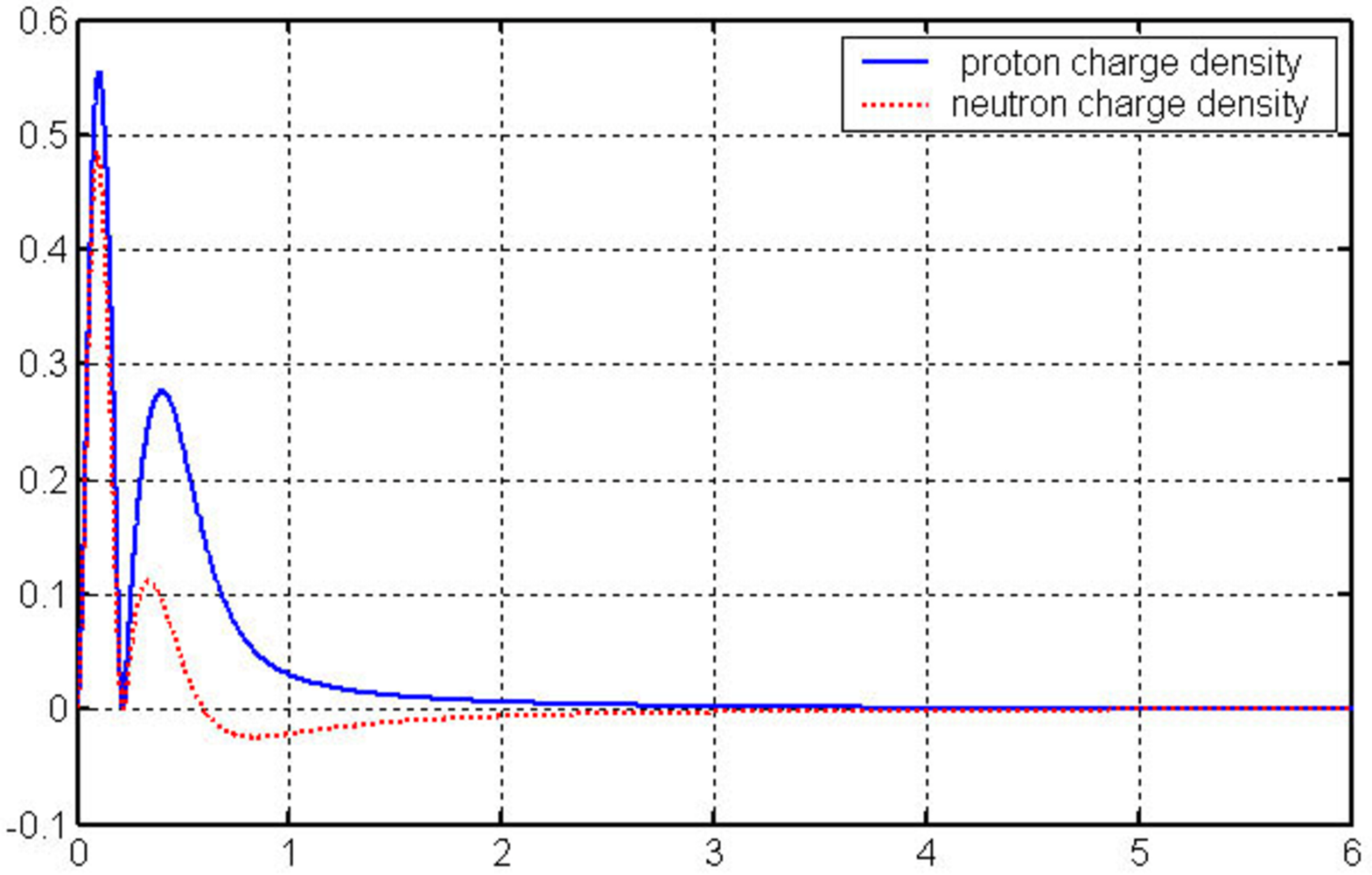}
\caption{Plots of density of neutron charge and proton charge for a case of $n=2$ [25].}
\end{figure}
\begin{figure}[htb]
\centering
\includegraphics[scale=0.40]{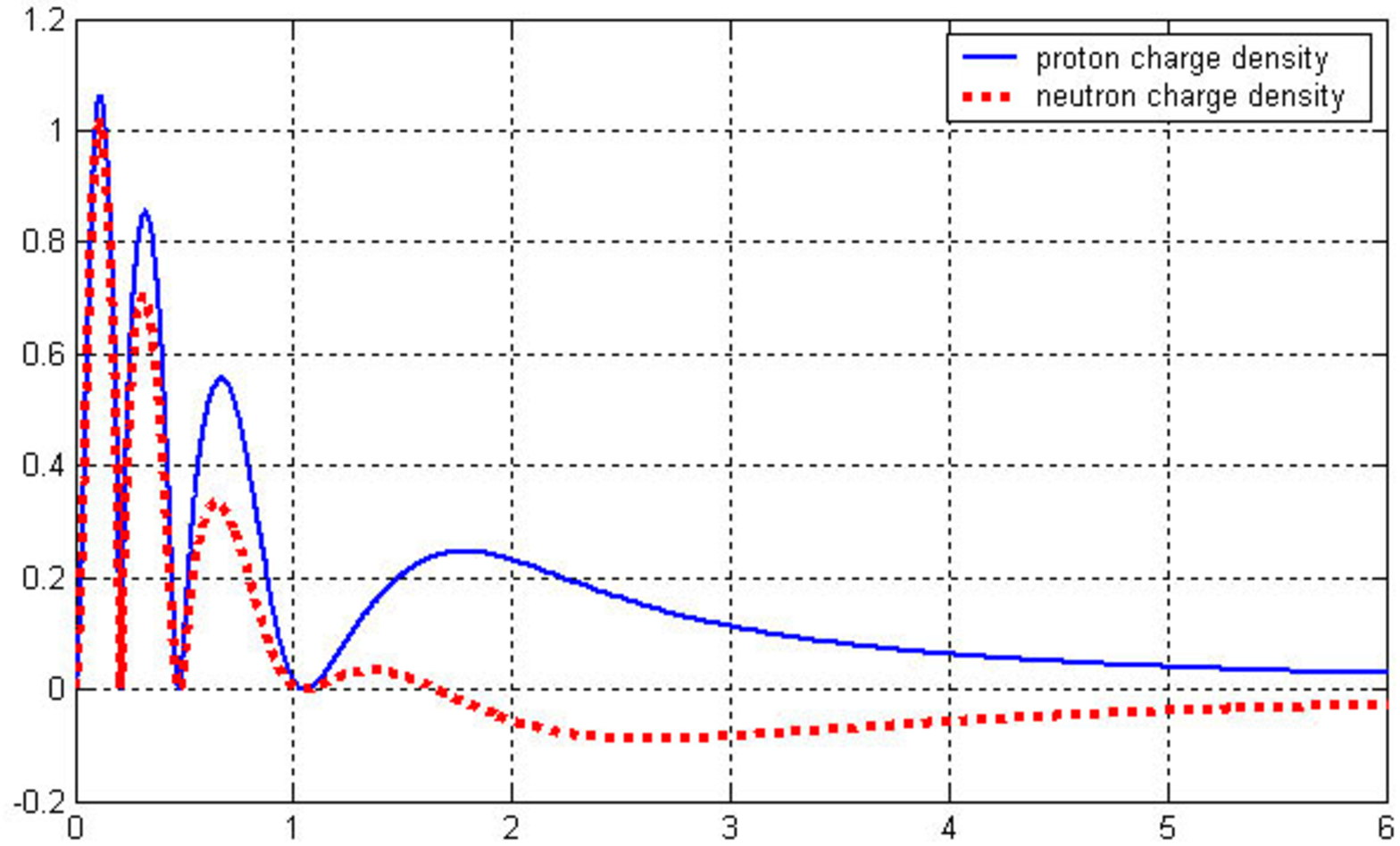}
\caption{Plots of density of neutron charge and proton charge for a case of $n=4$ [25].}
\end{figure}

\newpage
\centerline{\bf
NUMERICAL RESULTS TABLE FOR THE CASE OF N=4 [25]
}
\vskip0.4cm
\hspace{2cm} {\bf Quantity} \hspace{1.7cm} {\bf Prediction} \hspace{1.3cm} {\bf Experiment}
\vskip0.2cm
\hspace{2cm} $e$ \hspace{3.6cm} 1.10 \hspace{3.3cm} -
\vskip0.2cm
\hspace{2cm} $F_\pi $(MeV) \hspace{2.2cm} 168 \hspace{3.2cm} 186
\vskip0.2cm
\hspace{2cm} $\left\langle {r^2 } \right\rangle _{E,I = 0}^{1/2} $(fm) \hspace{1.5cm} 1.10 \hspace{3.1cm} 0.72
\vskip0.2cm
\hspace{2cm} $\left\langle {r^2 } \right\rangle _{M,I = 0}^{1/2} $(fm) \hspace{1.4cm} 1.93 \hspace{3.1cm} 0.81
\vskip0.2cm
\hspace{2cm} $\left\langle {r^2 } \right\rangle _{E,I = 1}^{1/2} $(fm) \hspace{1.6cm} $\infty $ \hspace{3.3cm} 0.88
\vskip0.2cm
\hspace{2cm} $\left\langle {r^2 } \right\rangle _{M,I = 1}^{1/2} $(fm) \hspace{1.6cm} $\infty $ \hspace{3.3cm} 0.8
\vskip0.2cm
\hspace{2cm} $\mu _{pro} $(mag) \hspace{2.1cm} 2.55 \hspace{3.1cm} 2.79
\vskip0.2cm
\hspace{2cm} $\mu _{neu} $(mag) \hspace{1.9cm} -0.66 \hspace{2.9cm} -1.91
\vskip0.2cm
\hspace{2cm} $\left| {\frac{{\mu _{pro} }}{{\mu _{neu} }}} \right|$ \hspace{2.9cm} 3.86 \hspace{3cm} 1.46
\vskip0.2cm
\hspace{2cm} $g_A $ \hspace{3.4cm} 3.52 \hspace{3cm} 1.23
\chapter{The composite Skyrme model with pion's mass}
\section{The term of pion'mass }
\vskip0.2cm

When Skyrme constructed a model from the theory of pi-meson, he did not let the term of mass of pion because of its spontaneous symmetry breaking. However, with $n = 1, 2, 3, 4$, datum of prediction are not really exact. One of causes that we do not let the term of pion's mass into the lagrangian [2, 3]. Therefore, we will propose the composite lagrangian density with the pion's mass [22, 24].
\vskip0.2cm
The lagrangian density now is
\begin{equation}
{\cal L}_n ^{pion}  = {\cal L}_n  + \frac{1}{{8n}}m_\pi ^2 F_\pi ^2 \left[ {Tr\left( {U^n } \right) - 2} \right] .
\end{equation}
\vskip0.2cm
Now using the hedgehog form gave by Skyrme $U_o \left( r \right) = \exp \left[ {i\tau .\hat rF\left( r \right)} \right]$, where $\tau $'s are Pauli's matrices and $\hat r = \frac{{\vec r}}{{\left| {\vec r} \right|}}$, the lagrangian density is
\begin{equation}
{\cal L}_n ^{pion}  = {\cal L}_n  + \frac{1}{{4n}}m_\pi ^2 F_\pi ^2 \left( {\cos nF - 1} \right) ,
\end{equation}
\begin{equation}
{\cal L}_n  = \frac{{F_\pi ^2 }}{8}\left[ {\left( {\frac{{dF}}{{dr}}} \right)^2  + 2\frac{{\sin ^2 nF}}{{n^2 r^2 }}} \right] + \frac{{\sin ^2 nF}}{{2e^2 n^2 r^2 }}\left[ {2\left( {\frac{{dF}}{{dr}}} \right)^2  + \frac{{\sin ^2 nF}}{{n^2 r^2 }}} \right] .
\end{equation}
\vskip0.2cm
The energy of hedgehog is
\begin{equation}
{\cal E}_n ^{pion}  = \frac{{F_\pi  }}{e}A_n  + \frac{1}{{e^3 F_\pi  }}\Theta _n ,
\end{equation}
with
\begin{equation}
A_n  = \int\limits_0^\infty  {4\pi \tilde r^2 \left\{ {\left[ {\frac{{F'^2 }}{8} + \frac{{\sin ^2 nF}}{{4n^2 \tilde r^2 }}} \right] + \frac{{e^2 F_\pi ^4 \sin ^2 nF}}{{n^2 \tilde r^2 }}\left[ {F'^2  + \frac{{\sin ^2 nF}}{{2n^2 \tilde r^2 }}} \right]} \right\}} d\tilde r ,
\end{equation}
and
\begin{equation}
\Theta _n  = \int\limits_0^\infty  {\frac{{\pi \tilde r^2 m_\pi ^2 }}{n}} \left( {\cos nF - 1} \right)d\tilde r ,
\end{equation}
$\tilde r = eF_\pi  r$ is a dimensionless variable.

\vskip0.2cm
After some steps of calculation as {\bf chapter 2}, we go to the hamiltonian
\begin{equation}
H_n  = {\cal E}_n ^{pion}  + \frac{1}{{8\Gamma _n }}\vec J^2  = {\cal E}_n ^{pion}  + \frac{1}{{8\Gamma _n }}\vec T^2 ,
\end{equation}
with $\vec J$ is a spin and $\vec T$ is an isospin. Eigenvalues of hamiltonian (3.7) are
\begin{equation}
M_n  = E_n  = \frac{{F_\pi  }}{e}A_n  + \frac{1}{{e^3 F_\pi  }}\Theta _n  + \frac{{l\left( {l + 2} \right)}}{{8\Gamma _n }} ,
\end{equation}
$l=2J$ and $J$ is the quantum number of spin.
\vskip0.2cm
So, the nucleon $\left( {J = 1/2} \right)$ and delta $\left( {J = 3/2} \right)$ masses are given by
\begin{equation}
M_N  = \frac{{F_\pi  }}{e}A_n  + \frac{1}{{e^3 F_\pi  }}\Theta _n  + \frac{{3e^3 F_\pi  }}{{8\Phi _n }} ,
\end{equation}
\begin{equation}
M_\Delta   = \frac{{F_\pi  }}{e}A_n  + \frac{1}{{e^3 F_\pi  }}\Theta _n  + \frac{{15e^3 F_\pi  }}{{8\Phi _n }} .
\end{equation}
\vskip0.2cm
From (3.9) and (3.10), we obtain
\begin{equation}
F_\pi   = \frac{{2\left( {M_\Delta - M_N } \right)\Phi _n }}{{3e^3 }} ,
\end{equation}
\begin{equation}
e = \sqrt[4]{{\frac{{4\left( {M_\Delta - M_N } \right)^3 \Phi _n^2 A_n }}{{6M_N \left( {M_\Delta - M_N } \right)^2 \Phi _n - 9\Theta _n - 3\left( {M_\Delta - M_N } \right)^2 \Phi _n }}}} .
\end{equation}
\section{The field equation and numerical results }
\vskip0.2cm
We obtain the nonlinear differential equation of $F\left( r \right)$ from minimum condition of the hedgehog's energy
\begin{equation}
\delta _F {\cal E}_n ^{pion} = 0 ,
\end{equation}
or
\begin{equation}
\frac{d}{{dr}}\frac{{\delta {\cal E}_n^{pion}}}{{\delta F' }} - \frac{{\delta {\cal E}_n^{pion}}}{{\delta F}} = 0 ( Euler - Lagrange - equation ) ,
\end{equation}
\begin{equation}
\left( {\frac{{\tilde r^2 }}{4} + \frac{2}{{n^2 }}\sin ^2 nF} \right)F'' + \frac{{\tilde r}}{2}F' + \frac{{\sin 2nF}}{n}\left( {F'^2  - \frac{{\sin ^2 nF}}{{n^2 \tilde r^2 }} - \frac{1}{4}} \right) - \beta \tilde r^2 \sin nF = 0 ,
\end{equation}
where
\begin{equation}
\beta  = \frac{{m_\pi ^2 }}{{4e^2 F_\pi ^2 }} .
\end{equation}
\vskip0.2cm

We see that (3.16) contains the constant of $e$ and $F_\pi $ but they are not known because they are must found after solving (3.15). Thus, we will solve (3.15) by the way is taking the value of $\beta $ in spite of $e$ and $F_\pi $ and raising it until it equals approximately the $\beta $ that defined by (3.11, 3.12, 3.16). In this case $n=4$, we need the value of $\beta $ is 0.0025 [24].
\begin{figure}[ht]
\centering
\includegraphics[scale=0.5]{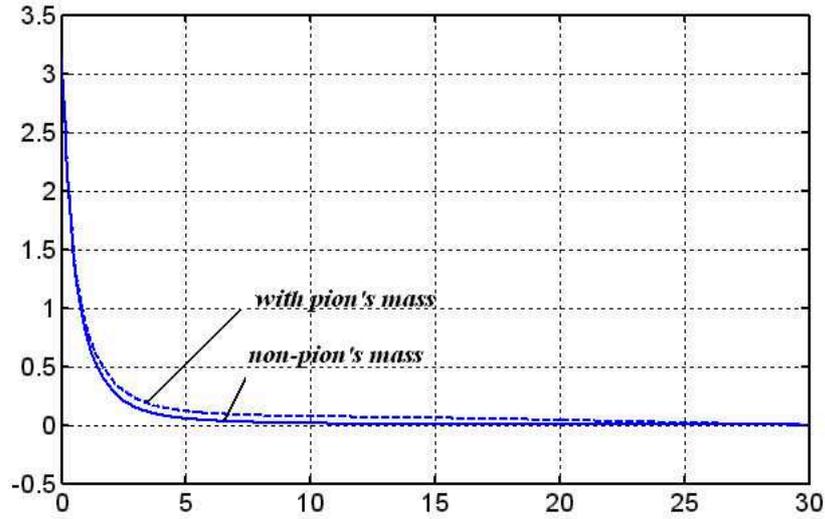}
\caption{Plots of numerical solution for a case of $n=4$ [24].}
\end{figure}

\begin{figure}[htb]
\centering
\includegraphics[scale=0.5]{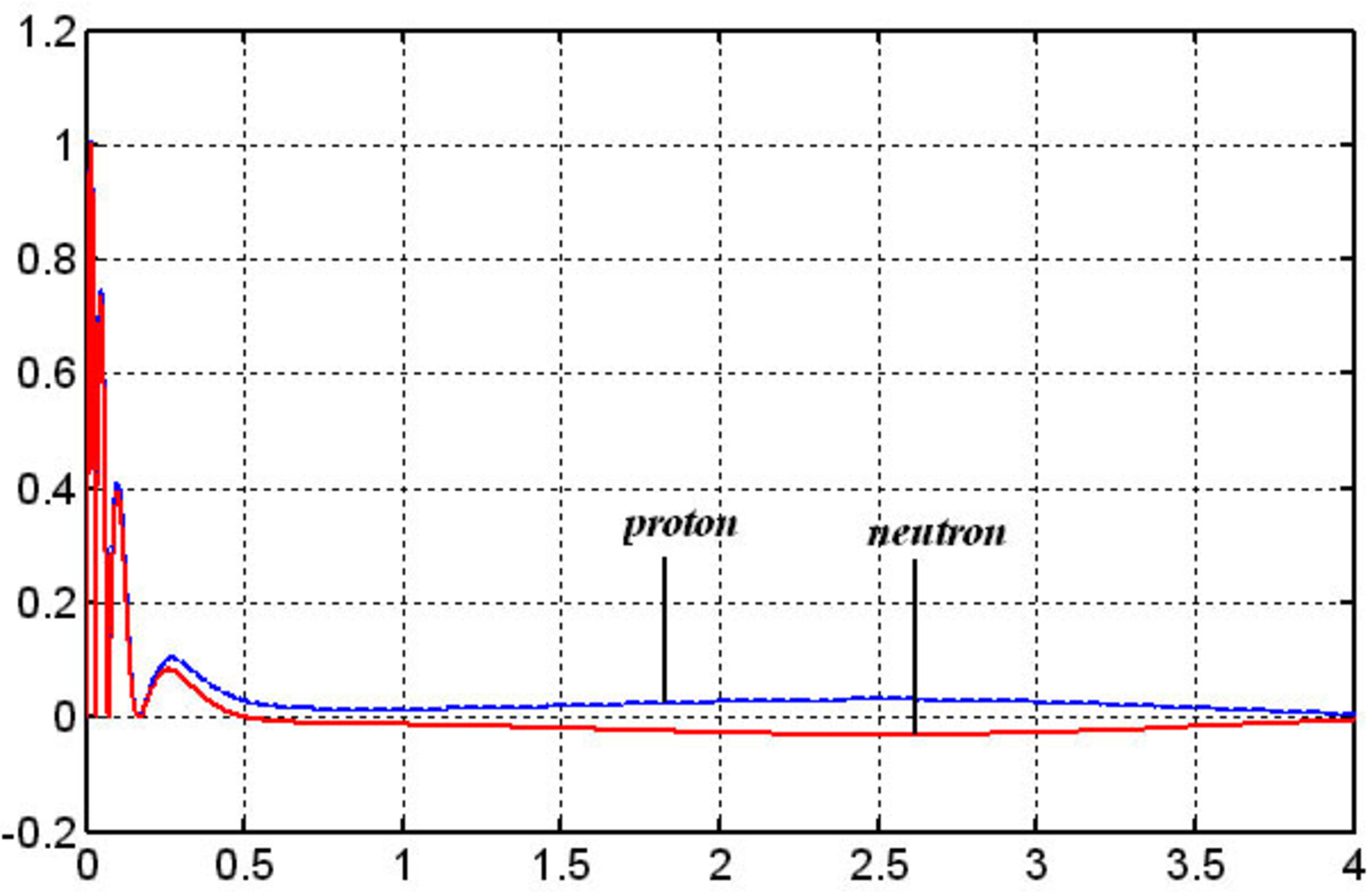}
\caption{Plots of density of neutron charge and proton charge for a case of $n=4$ [24].}
\end{figure}

\newpage
Following results that are obtained above by us ({\bf chapter 2}), quantities of baryon (proton and neutron) are described below [24]
\vskip0.4cm
\centerline{\bf
NUMERICAL RESULTS TABLE FOR THE CASE OF N=4
}
\vskip0.4cm
\hspace{2cm} {\bf Quantity} \hspace{1.7cm} {\bf Prediction} \hspace{1.2cm} {\bf Experiment}
\vskip0.2cm
\hspace{2cm} $e$ \hspace{3.8cm} 2.87 \hspace{3cm} -
\vskip0.2cm
\hspace{2cm} $F_\pi $(MeV) \hspace{2.2cm} 446.56 \hspace{2.5cm} 186
\vskip0.2cm
\hspace{2cm} $\left\langle {r^2 } \right\rangle _{E,I = 0}^{1/2} $(fm) \hspace{1.6cm} 0.21 \hspace{2.8cm} 0.71
\vskip0.2cm
\hspace{2cm} $\left\langle {r^2 } \right\rangle _{E,I = 1}^{1/2} $(fm) \hspace{1.5cm} 2.26 \hspace{2.9cm} 0.88
\vskip0.2cm
\hspace{2cm} $\left\langle {r^2 } \right\rangle _{M,I = 0}^{1/2} $(fm) \hspace{1.6cm} 1.0 \hspace{2.9cm} 0.81
\vskip0.2cm
\hspace{2cm} $\left\langle {r^2 } \right\rangle _{M,I = 1}^{1/2} $(fm) \hspace{1.5cm} 2.26 \hspace{2.8cm} 0.8
\vskip0.2cm
\hspace{2cm} $\mu _{pro} $(mag) \hspace{2.2cm} 1.64 \hspace{2.8cm} 2.79
\vskip0.2cm
\hspace{2cm} $\mu _{neu} $(mag) \hspace{2cm} -1.57 \hspace{2.6cm} -1.91
\vskip0.2cm
\hspace{2cm} $\left| {\frac{{\mu _{pro} }}{{\mu _{neu} }}} \right|$ \hspace{2.9cm} 1.04 \hspace{2.8cm} 1.46
\vskip0.2cm
\hspace{2cm} $g_A $ \hspace{3.45cm} 1.22 \hspace{2.8cm} 1.23
\leftskip0cm
\vskip0.4cm \vskip0.4cm
\chapter{The supersymmetric composite Skyrme model}

\vskip0.2cm

The fundamental theories of particles consider that elementary particles are points.
They all have the spin half-one and combining them together will
make particles which have the structural space. But, the model gave
by Skyrme is quite otherwise. He consider that there are solitons. In these models, one consider that soliton solutions have the structural space and the spin zero. By quantizing these solitons, we
obtain real particles that have the spin non-zero. Since working
with solitons, these models are all nonlinear. There is an important
point that these models work at the ideal case (a pion's mass is
zero). That makes datum of prediction are not really exact. Thus, we
need to let the term of pion's mass in these models. But, with only
pion, the models would not be exact, because we also have other
mesons. Must let all of mesons into the model. Thus, we extend the
model onto the \emph{SU(3)} group. But then will be an another cause
that makes the model is not really exact. In general we must
consider the contributions of the particle partners of the mesons.
On the other hand, each meson has a "superpartner" (make up from
quark superpartners). When all of these contributions are counted in
the model, it may be exact [23].
\vskip0.2cm

In 1984, E. A. Bergshoeff, R. I. Nepomachie, H. J. Schnitzer proposed Skyrmion of four-dimensional supersymmetric non-linear sigma model [5].
The composite Skyrme model proposed by H. Y. Cheung and F. Gursey in 1990. It's the general model in which the sigma model and $V-T$ model are two special cases corresponding $n=1,2$. Thus, we will calculate the composite model in the supersymmetry by the way proposed by E. A. Bergshoeff et al [26].
\vskip0.2cm

We recall the Lagrangian density of composite Skyrme model
\begin{equation}
{\cal L}_n  =  - \frac{{f_\pi ^2 }}{{16n^2 }}Tr\left( {\partial
_\mu  U^{ - n} \partial ^\mu  U^n } \right) + \frac{1}{{32e^2 n^4
}}Tr\left( {\left[ {U^{ - n} \partial _\mu  U^n ,U^{ - n} \partial
_\nu  U^n } \right]^2 } \right) ,
\end{equation}
with $f_\pi  $ is the pion decay constant, \emph{\textbf{e}} is a
dimensionless parameter.
\vskip0.2cm
Replacing ordinary derivatives in the Lagrangian density into covariant derivatives
\begin{equation}
D_\mu  U^n  = \partial _\mu  U^n  - iV_\mu  U^n \tau _3  ,
\end{equation}
the Lagrangian density (4.1) is now
\begin{equation}
{\cal L}_n  =  - \frac{{f_\pi ^2 }}{{16n^2 }}Tr\left( {D_\mu  U^{ - n} D^\mu  U^n } \right) + \frac{1}{{32e^2 n^4 }}Tr\left( {\left[ {U^{ - n} D_\mu  U^n ,U^{ - n} D_\nu  U^n } \right]^2 } \right) .
\end{equation}
and is invariant under local $U\left( 1 \right)_R $ and global $SU\left( 2 \right)_L $ transformations
\begin{equation}
U^n \left( r \right) \to AU^n \left( x \right)e^{i\lambda \left( r \right)\tau _3 },  A \in SU\left( 2 \right)_L ,
\end{equation}
\begin{equation}
V_\mu  \left( r \right) \to V_\mu  \left( r \right) + \partial _\mu  \lambda \left( r \right) .
\end{equation}
\vskip0.2cm
Really, we have
\begin{center}
$\left( {D_\mu  U^{ - n} } \right)' = \left[ {\partial _\mu   + i\left( {V_\mu   + \partial _\mu  \lambda } \right)\tau _3 } \right]\left( {U^{ - n} e^{ - i\lambda \left( x \right)\tau _3 } } \right)A^\dag $ ,
\end{center}
\begin{center}
$= \left( {\partial _\mu  U^{ - n} } \right)e^{ - i\lambda \tau _3 } A^\dag   - i\left( {\partial _\mu  \lambda } \right)\tau _3 U^{ - n} e^{ - i\lambda \tau _3 } A^\dag + $
\end{center}
\begin{center}
$+ iV_\mu  \tau _3 U^{ - n} e^{ - i\lambda \tau _3 } A^\dag   + i\left( {\partial _\mu  \lambda } \right)\tau _3 U^{ - n} e^{ - i\lambda \tau _3 } A^\dag $ ,
\end{center}
\begin{center}
$= \left[ {\left( {\partial _\mu   + iV_\mu  \tau _3 } \right)U^{ - n} } \right]e^{ -\lambda\left( x \right)\tau _3 } A^\dag $ ,
\end{center}
\begin{equation}
\bf{= \left( {D_\mu  U^{ - n} } \right)e^{ - i\lambda \left( x \right)\tau _3 } A^\dag}  ;
\end{equation}

\vskip0.2cm
\begin{center}
$\left( {D^\mu  U^n } \right)' = A\left[ {\partial ^\mu   - i\left( {V^\mu   + \partial ^\mu  \lambda } \right)\tau _3 } \right]\left( {U^n e^{i\lambda \left( x \right)\tau _3 } } \right)$,
\end{center}
\begin{center}
$ = A\left( {\partial ^\mu  U^n } \right)e^{i\lambda \tau _3 }  + iA\left( {\partial ^\mu  \lambda } \right)\tau _3 U^n e^{i\lambda \tau _3 }  - iAV^\mu  \tau _3 U^n e^{i\lambda \tau _3 }  - iA\left( {\partial ^\mu  \lambda } \right)\tau _3 U^n e^{i\lambda \tau _3 } $ ,
\end{center}
\begin{center}
$ = A\left( {\partial ^\mu  U^n } \right)e^{i\lambda \tau _3 }  - iAV^\mu  \tau _3 U^n e^{i\lambda \tau _3 }  = A\left[ {\left( {\partial ^\mu   - iV^\mu  \tau _3 } \right)U^n } \right]e^{i\lambda \tau _3 } $ ,
\end{center}
\begin{equation}
\bf{= A\left( {D^\mu  U^n } \right)e^{i\lambda \tau _3 }} ;
\end{equation}
\vskip0.2cm
\begin{center}
$ \Rightarrow \left( {D_\mu  U^{ - n} } \right)'\left( {D^\mu  U^n } \right)' = \left( {D_\mu  U^{ - n} } \right)e^{ - i\lambda \left( x \right)\tau _3 } A^\dag  Ae^{i\lambda \tau _3 } \left( {D^\mu  U^n } \right)$ ,
\end{center}
\begin{equation}
= \left( {D_\mu  U^{ - n} } \right)\left( {D^\mu  U^n } \right).
\end{equation}
\begin{equation}
\Rightarrow Tr\left[ {\left( {D_\mu  U^{ - n} } \right)'\left( {D^\mu  U^n } \right)'} \right] = Tr\left[ {\left( {D_\mu  U^{ - n} } \right)\left( {D^\mu  U^n } \right)} \right].
\end{equation}
\vskip0.2cm
The second term of Lagrangian is proved similarly.
\vskip0.2cm

The gauge field $V_\mu  \left( r \right)$ is defined
\begin{equation}
V_\mu   =  - \frac{i}{{2n}}Tr\left( {U^{ - n} \partial _\mu  U^n \tau _3 } \right) .
\end{equation}
\vskip0.2cm
We parameterize the $SU\left( 2 \right)$ matrix $U^n \left( r \right)$ in terms of the complex scalars $A_i $
\begin{equation}
U^n \left( r \right) = \left( {\begin{array}{*{20}c}
   {A_1 } & { - A_2^* }  \\
   {A_2 } & {A_1^* }  \\
\end{array}} \right) ,
\end{equation}
with the unitary constraint
\begin{equation}
U^\dag  U = 1 \Rightarrow \bar A^i A_i  = A_1^* A_1  + A_2^* A_2  = 1 .
\end{equation}
\vskip0.2cm
Now, (4.2) is
\begin{equation}
D_\mu  \left( {\begin{array}{*{20}c}
   {A_1 } & { - A_2^* }  \\
   {A_2 } & {A_1^* }  \\
\end{array}} \right) = \partial _\mu  \left( {\begin{array}{*{20}c}
   {A_1 } & { - A_2^* }  \\
   {A_2 } & {A_1^* }  \\
\end{array}} \right) - iV_\mu  \left( {\begin{array}{*{20}c}
   {A_1 } & {A_2^* }  \\
   {A_2 } & { - A_1^* }  \\
\end{array}} \right)
\end{equation}
or
\begin{equation}
D_\mu  A_i  = \left( {\partial _\mu   - iV_\mu  } \right)A_i ,
\end{equation}
\begin{equation}
D_\mu  \bar A_i  = \left( {\partial _\mu   + iV_\mu  } \right)\bar A_i .
\end{equation}
\vskip0.2cm
The form of gauge field is then
\begin{center}
$V_\mu   =  - \frac{i}{{2n}}Tr\left( {U^{ - n} \partial _\mu  U^n \tau _3 } \right)$ ,
\end{center}
\begin{center}
$ =  - \frac{i}{{2n}}Tr\left[ {\left( {\begin{array}{*{20}c}
   {A_1^* } & {A_2^* }  \\
   { - A_2 } & {A_1 }  \\
\end{array}} \right)\partial _\mu  \left( {\begin{array}{*{20}c}
   {A_1 } & { - A_2^* }  \\
   {A_2 } & {A_1^* }  \\
\end{array}} \right)\left( {\begin{array}{*{20}c}
   1 & 0  \\
   0 & { - 1}  \\
\end{array}} \right)} \right]$,
\end{center}
\begin{center}
$ =  - \frac{i}{{2n}}Tr\left[ {\left( {\begin{array}{*{20}c}
   {A_1^* } & {A_2^* }  \\
   { - A_2 } & {A_1 }  \\
\end{array}} \right)\left( {\begin{array}{*{20}c}
   {\partial _\mu  A_1 } & {\partial _\mu  A_2^* }  \\
   {\partial _\mu  A_2 } & { - \partial _\mu  A_1^* }  \\
\end{array}} \right)} \right]$,
\end{center}
\begin{center}
$ =  - \frac{i}{{2n}}Tr\left[ {\begin{array}{*{20}c}
   {A_1^* \partial _\mu  A_1  + A_2^* \partial _\mu  A_2 } & {}  \\
   {} & { - A_2 \partial _\mu  A_2^*  - A_1 \partial _\mu  A_1^* }  \\
\end{array}} \right]$ ,
\end{center}
\begin{center}
$ =  - \frac{i}{{2n}}\left[ {\bar A^i \partial _\mu  A_i  - \left( {\partial _\mu  \bar A^i } \right)A_i } \right]$ ,
\end{center}
\begin{equation}
\bf \Rightarrow{V_\mu  \left( r \right) =  - \frac{i}{{2n}}\bar A^i \mathord{\buildrel{\lower3pt\hbox{$\scriptscriptstyle\leftrightarrow$}}
\over \partial } A_i} .
\end{equation}
\vskip0.2cm
With matrix $U^n \left( r \right)$ is defined in (4.11) and the gauge field is defined in (4.16), the supersymmetric Lagrangian density (4.3) is then
\begin{equation}
{\cal L}_n  =  - \frac{{f_\pi ^2 }}{{8n^2 }}\bar D_\mu  \bar AD^\mu  A - \frac{1}{{16e^2 n^2 }}F_{\mu \nu }^2 \left( V \right) ,
\end{equation}
with
\begin{equation}
F_{\mu \nu } \left( V \right) = \partial _\mu  V_\nu   - \partial _\nu  V_\mu .
\end{equation}
\vskip0.2cm
Really, we have
\vskip0.2cm
+ {\bf The first term}
\begin{center}
${\cal L}_n ^1  =  - \frac{1}{{16n^2 }}f_\pi ^2 Tr\left[ {D^\mu  \left( {\begin{array}{*{20}c}
   {A_1^* } & {A_2^* }  \\
   { - A_2 } & {A_1 }  \\
\end{array}} \right)D_\mu  \left( {\begin{array}{*{20}c}
   {A_1 } & { - A_2^* }  \\
   {A_2 } & {A_1^* }  \\
\end{array}} \right)} \right]$,
\end{center}
\begin{center}
$ =  - \frac{1}{{16n^2 }}f_\pi ^2 Tr\left( {\begin{array}{*{20}c}
   {D^\mu  A_1^* D_\mu  A_1  + D^\mu  A_2^* D_\mu  A_2^* } & {}  \\
   {} & {D^\mu  A_2 D_\mu  A_2^*  + D^\mu  A_1 D_\mu  A_1^* }  \\
\end{array}} \right)$ ,
\end{center}
\begin{equation}
\Rightarrow{\bf{\cal L}_n ^1=  - \frac{1}{{8n^2 }}f_\pi ^2 D^\mu  \bar A^i D_\mu  A_i} .
\end{equation}
\vskip0.2cm

+ {\bf The second term}
\begin{center}
$U^{ - n} D_\mu  U^n U^{ - n} D_\nu  U^n  - U^{ - n} D_\nu  U^n U^{ - n} D_\mu  U^n $,
\end{center}
\begin{center}
$ = U^{ - n} \left[ {\left( {\partial _\mu   - iV_\mu  \tau _3 } \right)U^n } \right]U^{ - n} \left[ {\left( {\partial _\nu   - iV_\nu  \tau _3 } \right)U^n } \right]-$
\end{center}
\begin{center}
$ - U^{ - n} \left[ {\left( {\partial _\nu   - iV_\nu  \tau _3 } \right)U^n } \right]U^{ - n} \left[ {\left( {\partial _\mu   - iV_\mu  \tau _3 } \right)U^n } \right]$,
\end{center}
\begin{center}
$ = \left( {U^{ - n} \partial _\mu  U^n  - iU^{ - n} V_\mu  \tau _3 U^n } \right)\left( {U^{ - n} \partial _\nu  U^n  - iU^{ - n} V_\nu  \tau _3 U^n } \right)-$
\end{center}
\begin{center}
$ - \left( {U^{ - n} \partial _\nu  U^n  - iU^{ - n} V_\nu  \tau _3 U^n } \right)\left( {U^{ - n} \partial _\mu  U^n  - iU^{ - n} V_\mu  \tau _3 U^n } \right)$,
\end{center}
\begin{equation}
= U^{ - n} \partial _\mu  U^n U^{ - n} \partial _\nu  U^n  - U^{ - n} \partial _\nu  U^n U^{ - n} \partial _\mu  U^n  .
\end{equation}
\vskip0.2cm
Using
\begin{equation}
U^{ - n} U^n  = 1 \Rightarrow \partial _\mu  U^{ - n} U^n  + U^{ - n} \partial _\mu  U^n  = 0 ,
\end{equation}
\vskip0.2cm
(4.20) is now
\begin{equation}
\left( {4.20} \right) =  - \partial _\mu  U^{ - n} \partial _\nu  U^n  + \partial _\nu  U^{ - n} \partial _\mu  U^n  .
\end{equation}
\vskip0.2cm
The second term is now
\begin{equation}
\Rightarrow {\cal L}_n ^2  = \frac{1}{{32e^2 n^4 }}Tr\left[ {\left( {\partial _\mu  U^{ - n} \partial _\nu  U^n  - \partial _\nu  U^{ - n} \partial _\mu  U^n } \right)^2 } \right].
\end{equation}
\vskip0.2cm
We define
\begin{equation}
F_{\mu \nu } \left( V \right) \equiv \partial _\mu  V_\nu   - \partial _\nu  V_\mu ,
\end{equation}
\begin{equation}
 =  - \frac{i}{{2n}}Tr\left[ {\partial _\mu  \left( {U^{ - n} \partial _\nu  U^n \tau _3 } \right) - \partial _\nu  \left( {U^{ - n} \partial _\mu  U^n \tau _3 } \right)} \right],
\end{equation}
\begin{equation}
=  - \frac{i}{{2n}}Tr\left[ {\partial _\mu  U^{ - n} \partial _\nu  U^n \tau _3  - \partial _\nu  U^{ - n} \partial _\mu  U^n \tau _3 } \right].
\end{equation}
\begin{equation}
\Rightarrow F_{\mu \nu }^2  = \frac{1}{{2n^2 }}Tr\left[ {\left( {\partial _\mu  U^{ - n} \partial _\nu  U^n  - \partial _\nu  U^{ - n} \partial _\mu  U^n } \right)^2 } \right].
\end{equation}
\vskip0.2cm
From (4.23) and (4.27), we have
\begin{equation}
{\bf {\cal L}_n ^2  = \frac{1}{{16e^2 n^2 }}F_{\mu \nu }^2} .
\end{equation}
To supersymmetrise this model, we extend $A_i $ to chiral scalar multiples $\left( {A_i ,\psi _{\alpha i} ,F_i } \right)$ $\left( {i,\alpha  = 1,2} \right)$ and the vector $V_\mu  \left( r \right)$ to real vector multiples $\left( {V_\mu  ,\lambda _\alpha  ,D} \right)$. Here, the fields $F_i $ are complex scalars, $D$ is real scalar, $\psi _{\alpha i} $, $\lambda _\alpha  $ are Majorana two-component spinors. $\psi _{\alpha i} $ corresponds to a left-handed chiral spinor, $\bar \psi ^{\alpha i}  = \left( {\psi _i^\alpha  } \right)^* $  corresponds to a right-handed one. The supersymmetric Lagrangian density is given by
\begin{equation}
{\cal L}_{susy}  = \frac{{f_\pi ^2 }}{{8n^2 }}\left[ { - D^\mu  \bar A^i D_\mu  A_i  - \frac{1}{2}i\bar \psi ^{\dot \alpha i} \left( {\sigma _\mu  } \right)_{\alpha \dot \alpha } \mathord{\buildrel{\lower3pt\hbox{$\scriptscriptstyle\leftrightarrow$}}
\over D} ^\mu  \psi _i^\alpha   + \bar F^i F_i  - }\right.
\end{equation}
\begin{equation}
\left. { - i\bar A^i \lambda ^\alpha  \psi _{\alpha i}  + iA_i \bar \lambda ^{\dot \alpha } \bar \psi _{\dot \alpha }^i  + D\left( {\bar A^i A_i  - 1} \right)} \right] +
\end{equation}
\begin{equation}
+ \frac{1}{{8e^2 n^2 }}\left[ { - \frac{1}{2}F_{\mu \nu }^2  - i\bar \lambda ^{\dot \alpha } \left( {\sigma ^\mu  } \right)_{\dot \alpha }^\alpha  \partial _\mu  \lambda _\alpha   + D^2 } \right] ,
\end{equation}
is invariant under supersymmetric transformations
\begin{equation}
\delta A_i  =  - \varepsilon ^\alpha  \psi _{\alpha i} ,
\end{equation}
\begin{equation}
\delta \psi _{\alpha i}  =  - i\bar \varepsilon ^{\dot \alpha } \left( {\sigma ^\mu  } \right)_{\alpha \dot \alpha } D_\mu  A_i  + \varepsilon _\alpha  F_i ,
\end{equation}
\begin{equation}
\delta F_i  =  - i\bar \varepsilon ^{\dot \alpha } \left( {\sigma ^\mu  } \right)_{\dot \alpha }^\alpha  D_\mu  \psi _{\alpha i}  - i\bar \varepsilon ^{\dot \alpha } A_i \bar \lambda _{\dot \alpha } ,
\end{equation}
\begin{equation}
\delta V_\mu   =  - \frac{1}{2}i\left( {\sigma _\mu  } \right)^{\alpha \dot \alpha } \left( {\bar \varepsilon _{\dot \alpha } \lambda _\alpha   + \varepsilon _\alpha  \bar \lambda _{\dot \alpha } } \right) ,
\end{equation}
\begin{equation}
\delta \lambda _\alpha   = \varepsilon ^\beta  \left( {\sigma ^{\mu \nu } } \right)_{\beta \alpha } F_{\mu \nu }  + i\varepsilon _\alpha  D ,
\end{equation}
\begin{equation}
\delta D = \frac{1}{2}\left( {\sigma ^\mu  } \right)_{\alpha \dot \alpha } \partial _\mu  \left( {\bar \varepsilon ^{\dot \alpha } \lambda ^\alpha   - \varepsilon ^\alpha  \bar \lambda ^{\dot \alpha } } \right) .
\end{equation}
\vskip0.2cm
The field equation and their supersymmetric transformations lead to the following constraints
with
\begin{equation}
\bar A^i A_i  = 0 ,
\end{equation}
\begin{equation}
\bar A^i \psi _{\alpha i}  = 0 ,
\end{equation}
\begin{equation}
\bar A^i F_i  = 0
\end{equation}
and following algebraic expressions for
\begin{equation}
V_\mu   =  - \frac{1}{2}\left\{ {i\bar A^i \mathord{\buildrel{\lower3pt\hbox{$\scriptscriptstyle\leftrightarrow$}}
\over \partial } _\mu  A_i  + \left( {\sigma _\mu  } \right)^{\alpha \dot \alpha } \bar \psi _{\dot \alpha }^i \psi _{\alpha i} } \right\} ,
\end{equation}
\begin{equation}
\lambda _\alpha   =  - i\left\{ {\bar F^i \psi _{\alpha i}  + i\left( {\sigma ^\mu  } \right)_{\alpha \dot \alpha } \left( {D_\mu  A_i } \right)\bar \psi ^{\dot \alpha i} } \right\} ,
\end{equation}
\begin{equation}
D = D^\mu  \bar A^i D_\mu  A_i  + \frac{1}{2}i\bar \psi ^{\dot \alpha i} \left( {\sigma ^\mu  } \right)_{\alpha \dot \alpha } \left( {\mathord{\buildrel{\lower3pt\hbox{$\scriptscriptstyle\leftrightarrow$}}
\over D} _\mu  \psi _i^\alpha  } \right) - \bar F^i F_i .
\end{equation}
\vskip0.2cm
Setting $\psi _\alpha   = F_i  = 0$, then
\begin{equation}
{\cal L}_{susy}  =  - \frac{{f_\pi ^2 }}{{8n^2 }}\bar D^\mu  \bar AD_\mu  A + \frac{1}{{8e^2 n^2 }}\left[ { - \frac{1}{2}F_{\mu \nu }^2  + \left( {\bar D^\mu  \bar AD_\mu  A} \right)^2 } \right] .
\end{equation}
\vskip0.2cm
The second term is fourth-order in derivatives. However, there is other possible fourth-order term
\begin{equation}
\square \bar A\square A - \left( {\bar D^\mu  \bar AD_\mu  A} \right)^2 ,
\end{equation}
where $\square  \equiv D^\mu  D_\mu  $ is the gauge covariant d'Alembertian.
\vskip0.2cm
Thus, we may let it to the Lagrangian density
\begin{equation}
{\cal L}_{susy}  =  - \frac{{f_\pi ^2 }}{{8n^2 }}\bar D^\mu  \bar AD_\mu  A + \frac{1}{{8e^2 n^2 }}\left[ {\alpha \left\{ { - \frac{1}{2}F_{\mu \nu }^2  + \left( {\bar D^\mu  \bar AD_\mu  A} \right)^2 } \right\} + } \right.
\end{equation}
\begin{equation}
\left. { + \beta \left\{ {\square \bar A\square A - \left( {\bar D^\mu  \bar AD_\mu  A} \right)^2 } \right\}} \right] .
\end{equation}
\vskip0.2cm
Now, we recall the hedgehog ansatz
\begin{equation}
U^n \left( r \right) = \cos nf\left( r \right) + i\vec \tau \frac{{\vec r}}
{r}\sin nf\left( r \right) .
\end{equation}
\vskip0.2cm
From (4.11), we have
\begin{equation}
A_1  = \cos nf\left( r \right) + i\cos \theta \sin nf\left( r \right) ,
\end{equation}
\begin{equation}
A_2  = ie^{i\varphi } \sin \theta \sin nf\left( r \right) .
\end{equation}
\vskip0.2cm
Inserting the hedgehog ansatz into the supersymmetric Lagrangian density, we obtain the supersymmetric static energy
\begin{equation}
E = 4\pi \frac{{f_\pi  }}{e}\int {dxx^2 } \left\{ {\frac{1}{{12}}\left( {f'^2  + \frac{{2\sin ^2 nf}}{{n^2 x^2 }}} \right) + \left( {\frac{{\alpha  + \beta }}{{15}}} \right)\left( {f'^2  - \frac{{\sin ^2 nf}}{{n^2 x^2 }}} \right) + } \right.
\end{equation}
\begin{equation}
\left. { + \frac{\beta }{{12}}\left( {f'' + \frac{{2f'}}{{nx}} - \frac{{\sin 2nf}}{{n^2 x^2 }}} \right)} \right\} .
\end{equation}
\vskip0.2cm
The Euler-Lagrange equation takes the field equation
\begin{equation}
- x^2 f'' - 2xf' + \frac{{\sin 2nf}}{n} + \frac{{4\left( {\alpha  + \beta } \right)}}{5}\left[ {\frac{{2f''\sin ^2 nf}}{{n^2 }} - 6x^2 f'^2 f'' - 4xf'^3  + } \right.
\end{equation}
\begin{equation}
\left. { + \frac{{f'^2 \sin 2nf}}{n} + \frac{{\sin ^2 nf\sin 2nf}}{{n^3 x^2 }}} \right] + \beta \left[ {x^2 f^{\left( 4 \right)}  + \frac{{4xf^{\left( 3 \right)} }}{n} - \frac{{4f''\cos 2nf}}{n} + } \right.
\end{equation}
\begin{equation}
\left. { + \frac{{4f'^2 \sin 2nf}}{{n^2 }} - \frac{{4\sin ^2 nf\sin 2nf}}{{n^3 x^2 }}} \right] = 0 ,
\end{equation}
with $x= ef_\pi  r$ is the dimensionless variable and boundary conditions
\begin{equation}
f\left( 0 \right) = \pi,
\end{equation}
\begin{equation}
f\left( \infty  \right) = 0 .
\end{equation}

\begin{figure}[htb]
\centering
\includegraphics[scale=0.5]{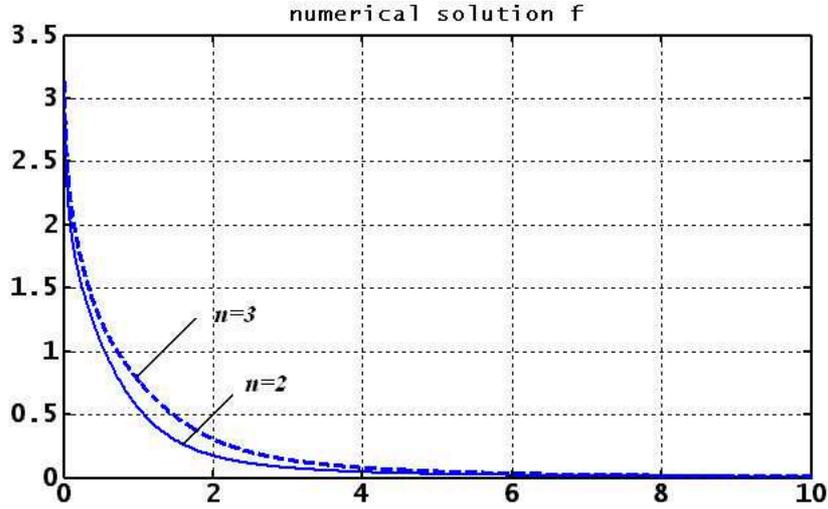}
\caption{Plots of supersymmetric numerical solution of $f\left( x \right)$ for a case of $\alpha  = 1,\beta  = 0,n = 2,3$ [26].}
\end{figure}

\leftskip0cm
\vskip0.4cm \vskip0.4cm
\chapter{Appendix}
\section{Some notations for Chapter 2 and 3}
There are some notations to calculations in chapter 2 and 3
\begin{equation}
\frac{\partial }{{\partial x^\mu  }} \equiv \partial _\mu   = \left( {\partial _0 ,\partial _i } \right) = \left( {\frac{\partial }{{\partial t}},\vec \nabla } \right) ;
\end{equation}
\begin{equation}
\frac{\partial }{{\partial x_\mu  }} \equiv \partial ^\mu   = \left( {\partial ^0 ,\partial ^i } \right) = g^{\mu \nu } \partial _\nu   = diag\left( {1, - 1, - 1, - 1} \right)\partial _\nu   = \left( {\frac{\partial }{{\partial t}}, - \vec \nabla } \right) ;
\end{equation}
\begin{equation}
\partial _i F\left( r \right) = \frac{{\partial F\left( r \right)}}{{\partial x^i }} = \frac{{\partial F}}{{\partial r}}\frac{{\partial r}}{{\partial x^i }} = F'\frac{{x^i }}{r} .
\end{equation}
\begin{equation}
U_o^n \left( r \right) = \exp \left[ {i\tau .\hat r\left\{ {nF\left( r \right)} \right\}} \right] = \cos \left\{ {nF\left( r \right)} \right\} + i\vec \tau .\hat r\sin \left\{ {nF\left( r \right)} \right\} ;
\end{equation}
\begin{equation}
U_o^{ - n} \left( r \right) = \exp \left[ { - i\tau .\hat r\left\{ {nF\left( r \right)} \right\}} \right] = \cos \left\{ {nF\left( r \right)} \right\} - i\vec \tau .\hat r\sin \left\{ {nF\left( r \right)} \right\} ;
\end{equation}
\vskip0.2cm
Some identities of Pauli matrices are
\begin{equation}
\tau _i \tau _j  = i\varepsilon _{ijk} \tau _k ;
\end{equation}
\begin{equation}
\left( {\tau _i } \right)^2  = 1 ;
\end{equation}
\begin{equation}
\tau _i \tau _j  =  - \tau _j \tau _i ;
\end{equation}
\begin{equation}
Tr\left( {\tau _i^2 } \right) = 2 ;
\end{equation}
\begin{equation}
Tr\left( {\tau _i } \right) = 0 ;
\end{equation}
\vskip0.2cm

We will define
\begin{center}
$\partial _\mu  U_o^n \left( r \right) = \partial _\mu  \left( {\cos nF} \right) + i\vec \tau \partial _\mu  \left( {\hat r\sin nF} \right)$ ,
\end{center}
\begin{center}
$= - n\partial _i F\sin nF + \frac{{i\vec \tau }}{r}\partial _i \vec r\sin nF + i\vec \tau \vec r\partial _i \left( {\frac{1}{r}} \right)\sin nF + in\vec \tau \frac{{\vec r}}{r}\partial _i F\cos nF $ ,
\end{center}
\begin{equation}
= - n\frac{{F'x^i }}{r}\sin nF + \frac{{i\tau _i \vec e^i }}{r}\sin nF - \frac{{i\vec \tau \vec rx^i }}{{r^3 }}\sin nF + in\frac{{\vec \tau \vec rF'x^i }}{{r^2 }}\cos nF ;
\end{equation}
\vspace{0.3cm}
\begin{center}
$\partial _\mu  U_o^{ - n} \left( r \right) = \partial _\mu  \left( {\cos nF} \right) - i\vec \tau \partial _\mu  \left( {\hat r\sin nF} \right)$ ,
\end{center}
\begin{equation}
= - n\frac{{F'x^i }}{r}\sin nF - \frac{{i\tau _i \vec e^i }}{r}\sin nF + \frac{{i\vec \tau \vec rx^i }}{{r^3 }}\sin nF - in\frac{{\vec \tau \vec rF'x^i }}{{r^2 }}\cos nF ;
\end{equation}
\vspace{0.3cm}
\begin{center}
$\partial ^\mu  U_o^n \left( r \right) = \partial ^\mu  \left( {\cos nF} \right) + i\vec \tau \partial ^\mu  \left( {\hat r\sin nF} \right)$ ,
\end{center}
\begin{equation}
= n\frac{{F'x_i }}{r}\sin nF - \frac{{i\tau _i \vec e_i }}{r}\sin nF + \frac{{i\vec \tau \vec rx_i }}{{r^3 }}\sin nF - in\frac{{\vec \tau \vec rF'x_i }}{{r^2 }}\cos nF ;
\end{equation}
\vskip0.2cm
The first term of the Lagrangian density (2.2) is defined
\begin{center}
$\Rightarrow {\cal L}_n ^1 = Tr\left( {\partial _\mu  U_o^{ - n} \partial ^\mu  U_o^n } \right)$
\end{center}
\begin{center}
$= Tr\left\{ {\left( { - n\frac{{F'x^i }}{r}\sin nF - \frac{{i\tau _i \vec e^i }}{r}\sin nF + \frac{{i\vec \tau \vec rx^i }}{{r^3 }}\sin nF - in\frac{{\vec \tau \vec rF'x_i }}{{r^2 }}\cos nF} \right) \times } \right.$
\end{center}
\begin{center}
$ \times \left. {\left( {n\frac{{F'x_i }}{r}\sin nF - \frac{{i\tau _i \vec e_i }}{r}\sin nF + \frac{{i\vec \tau \vec rx_i }}{{r^3 }}\sin nF - in\frac{{\vec \tau \vec rF'x_i }}{{r^2 }}\cos nF} \right)} \right\}$
\end{center}
\begin{center}
$= Tr\left\{ { - n^2 F'^2 \sin ^2 nF + inF'\frac{{\vec \tau \vec r}}{{r^2 }}\sin ^2 nF - inF'\frac{{\vec \tau \vec r}}{{r^2 }}\sin ^2 nF + } \right.$
\end{center}
\begin{center}
$ + in^2 F'^2 \frac{{\vec \tau \vec r}}{{2r}}\sin 2nF - inF'\frac{{\vec \tau \vec r}}{{r^2 }}\sin ^2 nF - 3\frac{{\sin ^2 nF}}{{r^2 }} + \frac{{\sin ^2 nF}}{{r^2 }} - $
\end{center}
\begin{center}
$ - n\frac{{F'}}{{2r}}\sin 2nF + inF'\frac{{\vec \tau \vec r}}{{r^2 }}\sin ^2 nF + \frac{{\sin ^2 nF}}{{r^2 }} - \frac{{\sin ^2 nF}}{{r^2 }} + n\frac{{F'}}{{2r}}\sin 2nF - $
\end{center}
\begin{center}
$\left. { - in^2 F'^2 \frac{{\vec \tau \vec r}}{{2r}}\sin 2nF - n\frac{{F'}}{{2r}}\sin 2nF + n\frac{{F'}}{{2r}}\sin 2nF - n^2 F'^2 \cos ^2 nF} \right\}$
\end{center}
\begin{center}
$ = Tr\left( { - n^2 F'^2  - 2\frac{{\sin ^2 nF}}{{r^2 }}} \right)$
\end{center}
\begin{equation}
\Rightarrow {\cal L}_n ^1 =  - 2n^2 \left( {\frac{{dF}}{{dr}}} \right)^2  - 4\frac{{\sin ^2 nF}}{{r^2 }}.
\end{equation}
\vskip0.2cm
To defined the second term of the Lagrangian density (2.2), we will go to the intermediary step using the unitary condition of $U_o^n$
\begin{equation}
U_o^{ - n} U_o^n  = 1 \Rightarrow \partial _\mu  U_o^{ - n} U_o^n  + U_o^{ - n} \partial _\mu  U_o^n  = 0 .
\end{equation}
\vskip0.2cm
The second term of Lagrangian density (2.2) is now
\begin{equation}
\Rightarrow {\cal L}_n ^2  = Tr\left[ {\left( {\partial _\mu  U_o^{ - n} \partial _\nu  U_o^n  - \partial _\nu  U_o^{ - n} \partial _\mu  U_o^n } \right)^2 } \right].
\end{equation}
\vskip0.2cm
Note (5.11), (5.12) we can define
\begin{equation}
{\cal L}_n ^2 = 32n^2 \frac{{\sin ^2 nF}}{{r^2 }}\left( {\frac{{dF}}{{dr}}} \right)^2  + 16\frac{{\sin ^4 nF}}{{r^4 }}.
\end{equation}

\section{Some notations for Chapter 4}
\vskip0.2cm
Dotted and undotted spinor indices run from 1 to 2 and are denoted by the early letters of the English alphabet. Spinor indices are raised and lowered by the $\varepsilon $ tensors
\begin{equation}
\varepsilon ^{ab}  = \left( {\begin{array}{*{20}c}
   0 & 1  \\
   { - 1} & 0  \\
\end{array}} \right),
\end{equation}
\begin{equation}
\varepsilon _{ab}  = \left( {\begin{array}{*{20}c}
   0 & { - 1}  \\
   1 & 0  \\
\end{array}} \right).
\end{equation}
\vskip0.2cm

For example
\begin{equation}
\psi ^a  = \varepsilon ^{ab} \psi _b ,
\end{equation}
\begin{equation}
\chi ^a  = \varepsilon _{ab} \chi ^b .
\end{equation}
\vskip0.2cm
The product $\psi \chi $ is
\begin{equation}
\psi \chi  = \psi ^a \chi _a  =  - \psi _a \chi ^a  = \chi ^a \psi _a  = \chi \psi .
\end{equation}
\vskip0.2cm
In the special case
\begin{equation}
\psi ^a \psi ^b  =  - \frac{1}{2}\psi \psi \varepsilon ^{ab} .
\end{equation}
\vskip0.2cm
The hermitian conjugates of the spinors $\psi$ and $\chi$ are $\bar \psi $ and $\bar\chi$
\begin{equation}
\bar \psi ^{\dot a}  = \left( {\psi ^a } \right)^\dag ,
\end{equation}
\begin{equation}
\bar \chi  = \left( {\chi _a } \right)^\dag .
\end{equation}
\vskip0.2cm
Their dotted indices are raised and lowered by the tensors $\varepsilon ^{\dot a\dot b} $ and $\varepsilon _{\dot a\dot b} $ which are equal to their undotted counterparts $\varepsilon ^{ab} $ and $\varepsilon _{ab} $. The product of them is
\begin{equation}
\bar \psi \bar \chi  = \bar \psi _{\dot a} \bar \chi ^{\dot a}  =  - \bar \psi ^{\dot a} \bar \chi _{\dot a}  = \bar \chi _{\dot a} \bar \psi ^{\dot a}  = \bar \chi \bar \psi .
\end{equation}
\vskip0.2cm
In the special case
\begin{equation}
\bar \psi ^{\dot a} \bar \psi ^{\dot b}  = \frac{1}{2}\bar \psi \bar \psi \varepsilon ^{\dot a\dot b} .
\end{equation}
\vskip0.2cm
The hermitian conjugates of the product $\chi \psi $ is
\begin{equation}
\left( {\chi \psi } \right)^\dag   = \left( {\chi ^a \psi _a } \right)^\dag   = \bar \psi _{\dot a} \bar \chi ^{\dot a}  = \bar \psi \bar \chi  = \bar \chi \bar \psi .
\end{equation}
\vskip0.2cm
The Pauli matrices $\sigma _{a\dot b}^m $ are
\begin{center}
$\sigma ^0  = \left( {\begin{array}{*{20}c}
   { 1} & 0  \\
   0 & { 1}  \\
\end{array}} \right),_{}^{} \sigma ^1  = \left( {\begin{array}{*{20}c}
   0 & 1  \\
   1 & 0  \\
\end{array}} \right)$ ,
\end{center}
\begin{equation}
\sigma ^2  = \left( {\begin{array}{*{20}c}
   0 & { - i}  \\
   i & 0  \\
\end{array}} \right),_{}^{} \sigma ^3  = \left( {\begin{array}{*{20}c}
   1 & 0  \\
   0 & { - 1}  \\
\end{array}} \right).
\end{equation}
\vskip0.2cm
Other useful identities are
\begin{equation}
\chi \sigma ^n \bar \psi  =  - \bar \psi \bar \sigma ^n \chi ,
\end{equation}
\begin{equation}
\chi \sigma ^m \bar \sigma ^n \psi  = \psi \sigma ^n \bar \sigma ^m \chi ,
\end{equation}
\begin{equation}
\left( {\chi \sigma ^m \bar \psi } \right)^\dag   = \psi \sigma ^m \bar \chi ,
\end{equation}
\begin{equation}
\left( {\chi \sigma ^m \bar \sigma ^n \psi } \right)^\dag   = \bar \psi \bar \sigma ^n \sigma ^m \bar \chi .
\end{equation}
\vskip0.2cm
Using these above notations, we can define (from...)
\begin{equation}
\delta \bar A_i  =  - \bar \psi _{\dot \alpha i} \bar \varepsilon ^{\dot \alpha } ,
\end{equation}
\begin{equation}
\delta \bar \psi _{\dot \alpha i}  = i\left( {\sigma ^\mu  } \right)_{\alpha \dot \alpha } \varepsilon ^\alpha  D_\mu  \bar A_i  + \bar \varepsilon _{\dot \alpha } \bar F_i ,
\end{equation}
\begin{equation}
\delta \bar F_i  = iD_\mu  \bar \psi _{\dot \alpha i} \left( {\sigma ^\mu  } \right)_{\dot \alpha }^\alpha  \varepsilon ^\alpha   + i\bar A_i \lambda _\alpha  \varepsilon ^\alpha ,
\end{equation}
\begin{equation}
\delta \bar V_\mu   = \frac{1}{2}i\left( {\bar \lambda _{\dot \alpha } \varepsilon _\alpha   + \lambda _\alpha  \bar \varepsilon _{\dot \alpha } } \right)\left( {\sigma _\mu  } \right)^{\alpha \dot \alpha } ,
\end{equation}
\begin{equation}
\delta \bar \lambda _{\dot \alpha }  = \left( {\sigma ^{\mu \nu } } \right)_{\beta \alpha } \bar \varepsilon ^{\dot \beta } F_{\mu \nu }  - i\bar \varepsilon _{\dot \alpha } \bar D ,
\end{equation}
\begin{equation}
\delta \bar D = \frac{1}{2}\partial _\mu  \left( {\bar \lambda ^{\dot \alpha } \varepsilon ^\alpha   - \lambda ^\alpha  \bar \varepsilon ^{\dot \alpha } } \right)\left( {\sigma ^\mu  } \right)_{\alpha \dot \alpha } .
\end{equation}
\vskip0.2cm
Since $\delta $ and $D_\mu  $ can permute together, $\delta {\cal L}_{susy} $ is defined
\begin{center}
$\delta {\cal L}_{susy}  = \frac{{f_\pi ^2 }}{{8n^2 }}\left[ { - D^\mu  \left( {\delta \bar A^i } \right)D_\mu  A_i  - D^\mu  \bar A^i D_\mu  \left( {\delta A_i } \right) - \frac{1}{2}i\left( {\delta \bar \psi ^{\dot \alpha i} } \right)\left( {\sigma _\mu  } \right)_{\alpha \dot \alpha } \mathord{\buildrel{\lower3pt\hbox{$\scriptscriptstyle\leftrightarrow$}}
\over D} ^\mu  \psi _i^\alpha   - } \right.$
\end{center}
\begin{center}
$ - \frac{1}{2}i\bar \psi ^{\dot \alpha i} \left( {\sigma _\mu  } \right)_{\alpha \dot \alpha } \mathord{\buildrel{\lower3pt\hbox{$\scriptscriptstyle\leftrightarrow$}}
\over D} ^\mu  \left( {\delta \psi _i^\alpha  } \right) + \delta \bar F^i F_i  + \bar F^i \delta F_i  - i\left( {\delta \bar A^i } \right)\lambda ^\alpha  \psi _{\alpha i}  - i\bar A^i \left( {\delta \lambda ^\alpha  } \right)\psi _{\alpha i}  - i\bar A^i \lambda ^\alpha  \delta \psi _{\alpha i}  + $
\end{center}
\begin{center}
$\left. { + i\left( {\delta A_i } \right)\bar \lambda ^{\dot \alpha } \bar \psi _{\dot \alpha }^i  + iA_i \left( {\delta \bar \lambda ^{\dot \alpha } } \right)\bar \psi _{\dot \alpha }^i  + iA_i \bar \lambda ^{\dot \alpha } \left( {\delta \bar \psi _{\dot \alpha }^i } \right) + \left( {\delta D} \right)\left( {\bar A^i A_i  - 1} \right) + D\left( {\delta \bar A^i } \right)A_i  + D\bar A^i \left( {\delta A_i } \right)} \right] + $
\end{center}
\begin{equation}
+ \frac{1}{{8e^2 n^4 }}\left[ { - F_{\mu \nu }^{} \delta F_{\mu \nu }^{}  - i\left( {\delta \bar \lambda ^{\dot \alpha } } \right)\left( {\sigma ^\mu  } \right)_{\dot \alpha }^\alpha  \partial _\mu  \lambda _\alpha   - i\bar \lambda ^{\dot \alpha } \left( {\sigma ^\mu  } \right)_{\dot \alpha }^\alpha  \partial _\mu  \left( {\delta \lambda _\alpha  } \right) + 2D\delta D} \right].
\end{equation}

\end{document}

%% file: seteps.tex
\def\centereps#1#2#3{\vskip#2\relax\centerline{\hbox to#1{\special
  {eps:#3 x=#1, y=#2}\hfil}}}